\documentclass[prd,aps,10pt,twocolumn,floatfix,nofootinbib,superscriptaddress]{revtex4-1}
\pdfoutput=1
\usepackage{lipsum}
\usepackage{textcomp}
\usepackage{amsmath}
\usepackage{amsfonts}
\usepackage{amssymb}
\usepackage{graphicx}
\usepackage{hyperref}
\usepackage{subfigure}
\usepackage[usenames, dvipsnames]{color}

\begin{document}

\title{Pulsar magnetospheres in General Relativity} 

\author{Federico Carrasco}
\email{fcarrasc@famaf.unc.edu.ar}
\affiliation{Facultad de Matem\'{a}tica, Astronom\'{i}a, F\'{i}sica y Computaci\'{o}n, Universidad Nacional de C\'o{}rdoba.\\
Instituto de F\'{i}sica Enrique Gaviola, CONICET. Ciudad Universitaria (5000), C\'o{}rdoba, Argentina. }
\affiliation{Departament de F\'i{}sica \& IAC3, Universitat de les Illes Balears and Institut d'Estudis Espacials de Catalunya, Palma de Mallorca, Baleares E-07122, Spain.}
\author{Carlos Palenzuela}
\email{carlos.palenzuela@uib.es }
\affiliation{Departament de F\'i{}sica \& IAC3, Universitat de les Illes Balears and Institut d'Estudis Espacials de Catalunya, Palma de Mallorca, Baleares E-07122, Spain.}
\author{Oscar Reula}
\email{reula@famaf.unc.edu.ar}
\affiliation{Facultad de Matem\'{a}tica, Astronom\'{i}a, F\'{i}sica y Computaci\'{o}n, Universidad Nacional de C\'o{}rdoba.\\
Instituto de F\'{i}sica Enrique Gaviola, CONICET. Ciudad Universitaria (5000), C\'o{}rdoba, Argentina. }

\date{\today}

\begin{abstract}
The main contribution to the pulsar power can be calculated by assuming a rotating magnetically-dominated magnetosphere described by the force-free approximation. Although this simple model has been used thoroughly to study pulsar magnetospheres in the flat spacetime regime, only few works have considered the relativistic corrections introduced by the curvature and frame-dragging effects induced by a rotating neutron star. Here we revisit the problem and describe pulsar magnetospheres within full General Relativity, quantifying the corrections as a function of the angular velocity, the compactness of the star and the misalignment angle between the spin and the magnetic dipole. We provide analytical expressions for the pulsar luminosity by fitting our numerical results. Finally, we also analyze the effect of the relativistic corrections on the braking index, which indicates a slight increment in its value. 
\end{abstract}


\maketitle

\section{Introduction}

Pulsars are bright sources of electromagnetic radiation, emitting from radio to gamma-ray frequencies. Even though the main picture --consisting on a rotating magnetized neutron star-- is rather simple,
a full solution that explain the diverse observational phenomenology still remains elusive due to the extreme conditions found on neutron stars. With typical radius $R=10-13$ km and stellar masses between $M\approx1.2-2.0\, M_{\odot}$\footnote{www3.mpifr-bonn.mpg.de/staff/pfreire/NS\_masses.html}, they are very compact astrophysical objects, only exceeded by black holes. Moreover, 
their rotation periods ranges from seconds to millisecond, and their surface magnetic field intensities, inferred by timing properties, vary from $\sim 10^8$ to $\sim 10^{15}$ G (see McGill and ATNF pulsar catalogs  \footnote{www.physics.mcgill.ca/$\sim$pulsar/magnetar/main.html}  \footnote{www.atnf.csiro.au/people/pulsar/psrcat/}).

In a seminal work, Goldreich \& Julian \cite{goldreich} first demonstrated that the neutron star rotation induces electric fields which are strong enough to strip charged particles off the stellar surface. This mechanism and the pair production induced by the interaction of high-energy photons eventually populate the surrounding magnetosphere with a tenuous plasma.
In such rarefied environments, the electromagnetic force dominates over particle inertia and leads to the force-free assumption, which is thought to hold everywhere except in small regions of space called gaps, where the observed radiation is produced by means of particle acceleration.
This low-inertia limit of relativistic magnetohydrodynamics (MHD), known as force-free electrodynamics (FFE), simplifies the problem considerably and  
has been widely used to study the global properties of pulsar magnetospheres. 
Assuming a dipolar magnetic field aligned with the rotation axis of the neutron star a canonical model emerged, beginning with a solution found by Contopoulos, Kazanas \& Fendt \cite{contopoulos1999} (CKF hereafter).
The CKF solution confirmed the basic qualitative picture originally sketched by Goldrich \& Julian;
namely, the existence of a closed zone that co-rotates with the star, together with a polar outflow along open magnetic field lines that extends to infinity. 
These two regions are separated by thin return current layers emanating from the poles and meeting at the Y-point, where the light cylinder intersects the equatorial plane. Beyond the Y-point there is a strong current sheet which extends along the equator up to infinity.
After the CKF solution first appearance, other steady solutions were found~\cite{goodwin2004, gruzinov2006, timokhin2006force}, 
displaying a Y-point located at different positions within the light cylinder and thus raising the question about uniqueness of the CKF solution.
Time-dependent force-free simulations later revealed that axi-symmetric solutions indeed converge to a CKF-type configuration. 
These simulations have been developed an extended by several authors using diverse numerical techniques. Both aligned \cite{mckinney2006relativistic, timokhin2006force, contopoulos2006, yu2011, PHAEDRA} and 
oblique \cite{spitkovsky2006, kalapotharakos2009, petri2012} pulsar magnetospheres were widely studied on flat spacetime.
The effects of plasma pressure and inertia were further included (e.g.~\cite{komissarov2006simulations, gruzinov2008diss, li2012, kalapotharakos2012, tchekhovskoy2013, cao2016});
and the assumption of a dipolar magnetic field has been recently relaxed \cite{rezzolla2004electromagnetic, bonazzola2015general, petri2015multipolar, petri2017multipolar} 
by considering more general multipolar field configurations in electrovacuum.
On the other hand, works including curvature effects due to general relativity (GR) are somewhat scarce.
The first time-dependent simulations incorporating GR effects appeared in \cite{Palenzuela2013}, 
reporting an approximate $20\% $ enhancement of the aligned rotator spin-down luminosity relative to its flat spacetime value.
Later, Ruiz et al. \cite{ruiz2014} presented a more systematic study of the GR aligned rotator, 
providing fitting formulas that shows the luminosity dependence on the star rotation and compactness. 
These works, although using different numerical codes, were both performed by using finite-difference schemes and matching the stellar ideal MHD interior to a force-free exterior. 
Recently, Petri \cite{petri2015general} has employed a pseudo-spectral code with discontinuous Galerkin methods 
to evolve three-dimensional pulsar magnetospheres, modeling the NS by suitable boundary conditions at the stellar surface. 
This work thoroughly analyzes the spin-down luminosity dependence on the rotation rate and 
the misalignment angle between the spin and the magnetic dipolar moment, comparing Newtonian and GR results at a fixed compactness.
More recently, there has been also an interesting analytical approach to the slowly rotating limit \cite{gralla2016pulsar, gralla2017inclined}, 
using some tools from differential geometry such as the exterior calculus.

In the last years there have been significant progress on particle-in-cell (PIC) simulations, which had allowed to reproduce global properties of pulsar magnetospheres, including self-consistently the regions of plasma production and acceleration (see e.g.~\cite{philippov2014ab, chen2014, philippov2015oblique}). 
One of the general lessons emerging from these PIC simulations is that pulsar environments are nearly force-free everywhere except for the thin return current layers and the current sheet outside the light cylinder,  
where particles are produced and part of the radiation is generated.
Moreover, luminosities obtained within the PIC framework are consistent with previous force-free simulations; 
even for the oblique rotator case where the $1+\sin^2 \chi$ dependence on the inclination angle $\chi$ has been retrieved \cite{philippov2014}.
Notice however that the inclusion of curvature --and in particular, the frame dragging effects-- has shown to be crucial in allowing pair formation near the polar cap regions \cite{philippov2015ab}.

In this paper we aim to elucidate the role of relativistic and curvature effects on the spin-down luminosity of a pulsar, by performing full 3D time-dependent simulations of force-free pulsar magnetospheres within general relativity. The presence of the rotating neutron star is modeled by means of suitable boundary conditions at the stellar surface, while the exterior spacetime is approximated by the Kerr metric. It has been further assumed that the star posses a dipolar magnetic field as in most previous studies. 
Our numerical code, based on high-order finite differences schemes over a multiple patch infrastructure \cite{Leco_1}, 
has been extended from previous efforts in the context of black holes magnetospheres \cite{FFE2},  e.g.~\cite{Leco_1}).
The numerical domain is equipped with a Kerr metric (in appropriate coordinates) and accommodates quite naturally to the geometry of the problem.
This code has been further developed in the present work, by including novel boundary conditions to represent the perfectly conducting surface of the NS.
The method to deal with such boundary conditions relies on the \textit{penalty technique} \cite{Carpenter1994, Carpenter1999, Carpenter2001},  
and uses the characteristic decomposition of the force-free equations employed for the evolution. 
Although this numerical implementation is very different from those of previous studies like Ruiz et al. \cite{ruiz2014} and Petri \cite{petri2015general},
the results found on this paper are in good agreement in the regimes on which they overlap. As a final result, we provide a general expression describing the spin-down luminosity in terms of the three adimensional parameters that specify the pulsar: the spin rate of the neutron star, the stellar compactness and the misalignment angle between spin and magnetic dipole axis. 

This article is organized as follows: In Section II we present the main aspects of our numerical approach and setup. 
Especial attention has been devoted to the treatment of boundary conditions at the stellar surface, 
while some related technical details --as well as other examples of application-- were deferred to Appendix A. 
The main results are presented in Section III, starting with some tests that shows the correct implementation of boundary conditions and constraints behavior.
We compare our numerical results with previous studies and then propose a generic formula for the pulsar spin-down luminosity, deduced from fittings of the numerical data.
Conclusions and some perspectives for future projects are presented on Section IV. 
Throughout, we adopt geometrized units in which $c=G=1$ and Lorentz-Heaviside units for the electromagnetic field.

\section{Setup}

Here we are interested on modeling the magnetospheres of neutron stars by
solving a particular version of force-free electrodynamics obtained at \cite{FFE},
which has some improved properties in terms of well posedness and involves the full force-free current density~\footnote{Similar hyperbolic formulations were presented in ref.'s~\cite{pfeiffer,Pfeiffer2015}}. 
The numerical scheme to solve these equations is based on the \textit{multi-block approach} \cite{Leco_1, Carpenter1994, Carpenter1999, Carpenter2001}, 
where a particular multiple patch infrastructure has been equipped with the Kerr metric \cite{Leco_1}. 
This provides a domain perfectly adapted to the geometry of the problem, having two global (inner and outer) boundaries with spherical topology. 
This implementation has been employed to perform accurate studies of force-free jets in black hole spacetimes~\cite{FFE2}. 
Here we introduce however an important modification to the previous scheme: we develop detailed boundary conditions for the inner surface (placed inside the black hole in \cite{FFE2}), as to represent the perfect conducting surface of the star. 

We shall start this section by briefly summarizing the generic features of the numerical approach. Then, we introduce the set of equations used to evolve the system. Finally, we describe how to deal with the boundary conditions for the stellar surface through the \textit{penalty method} and the initial data
for the magnetic field.\\

\subsection{General Scheme}

Our numerical domain consists on several touching grids (i.e.~there is no overlap among them and only points at the boundaries are sheared), 
commonly referred as \textit{multi-block approach} \cite{Leco_1}. 
%
%
The equations are discretized at each individual subdomain by using difference operators constructed to satisfy summation by
parts (SBP). 
In particular, we employ sixth-order accurate difference operators on the interior and third-order at the boundaries.
\textit{Penalty terms} \cite{Carpenter1994, Carpenter1999, Carpenter2001} are added to the evolution equations at boundary points.
These terms penalize possible mismatches between the different values the characteristic fields take at the interfaces, 
providing a consistent way of communicate information between the different blocks.
More concretely, the penalization terms at the boundary points of a subdomain (say, ``A'') modify the evolution equations as, 
 \begin{equation}\label{interfaces}
  \partial_t U_{A}^{\mu} \rightarrow  \partial_t U_{A}^{\mu} + \frac{1}{ h \, \sigma_{o}} \sum_{a (\lambda_a >0)} \lambda_{a} \,  P^{\mu}_{(a)\nu} \left( U^{\nu}_B - U^{\nu}_A \right) 
 \end{equation}
where $U^{\nu}_B$ are the values of the fields on the overlapping points from a neighbor subdomain ``B'' and the index ``$a$''  labels the different characteristic modes, being  $\lambda_{a}$ their  eigenvalues and $P^{\mu}_{(a) \nu} (\cdot)$ the projectors into their characteristic subspaces. 
Summation is performed only over the incoming  modes (respect to subdomain A), $\lambda_{a} > 0$.
While $h$ is the grid spacing along the normal direction and $\sigma_{o}$ the coefficient defining a discrete scalar product (see eqn.~\eqref{semi-discrete} below)
valuated at the boundary. 
The projectors are build from the eigen-basis $\left\lbrace U^{\mu}_{(a)} \right\rbrace$ (and co-basis $\left\lbrace \Theta_{\mu}^{(a)} \right\rbrace$) as,
\begin{equation*}
 P^{\mu}_{(a)\nu} := U_{(a)}^{\mu} \, \Theta^{(a)}_{\nu}  
\end{equation*}
For each incoming mode at one side of the interface there is an associated outgoing mode 
from the other side, and the penalty essentially enforces those values to coincide. 

At each subdomain, it is possible to find a semidiscrete energy defined by both a symmetrizer of the system at the continuum and a discrete scalar product with respect to which SBP holds,
\begin{equation}\label{semi-discrete}
 \textlangle u, v \textrangle  := h_{x} \, h_{y} \, h_{z} \sum_{i,j,k} (u_{i,j,k}, v_{i,j,k}) \, \sigma^{x}_i \, \sigma^{y}_j \, \sigma^{z}_k
\end{equation}
The summation by parts property of the operators implies an energy estimate, up to outer boundary and interface terms.
The penalties are constructed such that their contribution to the energy estimate cancel inconvenient interface terms, thus providing an energy estimate which covers the whole integration region across grids. 
Such semidiscrete energy estimates guarantee the stability of the numerical scheme, provided an appropriate time integrator is chosen \cite{Leco_2}.
A fourth order Runge-Kutta algorithm is used for time integration in our code. 
We incorporate numerical dissipation through adapted Kreiss-Oliger operators \cite{Tiglio2007}, which are eight-order accurate on the interior and fourth-order at the boundaries.

Each subdomain is handed to a separate processor, while the information required for the interfaces treatment is communicated among them by the \textit{message passing interface} (MPI) system.
The computation for each grid may be, as well, parallelized by means of OpenMP. \break

\subsection{Evolution Equations}

We shall start from the covariant version of force-free electrodynamics for the electromagnetic field $F_{ab}$ and the Faraday tensor 
$F^{*}_{ab} = \epsilon_{abcd} F^{cd}/2$, 
as presented in ref.~\cite{FFE}:
\begin{eqnarray}
 \tilde{F}^{ab} \nabla^c F_{bc} &=& 0 \label{forcefree1} \\
 \nabla_b F^{*ab} + \nabla^a \phi &=& \kappa n^a \phi \label{forcefree2} \\
 \tilde{F}^{*bc} \nabla^a F_{bc} &=& 0 \label{forcefree3}
\end{eqnarray}
where the field 
\begin{equation}\label{F_tilde}
\tilde{F}_{ab} := F_{ab} + \sigma F^{*}_{ab} \quad \text{ ; } \quad \sigma = \frac{G}{F+\sqrt{F^2 + G^2}}
\end{equation}
was defined to extend the system outside of the constraint submanifold, $G = 0$, being $G:= F^{ab}F^{*}_{ab}$ and $F:= F^{ab}F_{ab}$ the two electromagnetic invariants. Notice that eq.~(\ref{forcefree1}) reduces 
to the force-free condition, eq.~(\ref{forcefree2}) is the Faraday equation with an extra field $\phi$ to dynamically control the magnetic divergence-free constraint (see e.g.~\cite{Dedner, Komissarov2004b, Mari}), and eq.~(\ref{forcefree3}) is just the generalization of the constraint condition $\nabla_a G =0$.

A covariant hyperbolitation for these equations was found in \cite{FFE}, ensuring well posedness of the symmetrized system (equations (32)-(33) in \cite{FFE}). 
The final evolution equations are then obtained after performing a standard 3+1 decomposition of such system. 
The line element in the adapted coordinates of the foliation is given by,
\begin{equation*}
ds^2 = (\beta^2 - \alpha^{2}) dt^2 + 2 \beta_i dx^i dt +\gamma_{ij} dx^i dx^j
\end{equation*}
with $\alpha$,$\beta^i$ and $\gamma_{ij}$ being the \textit{lapse function}, the \textit{shift vector} and the \textit{intrinsic metric} on the spatial slices, respectively.

%
We follow the conventions adopted in e.g.~\cite{palenzuela2010}, where the electric and magnetic field are taken to be $E_a := F_{ab} n^b$ and $B_a := F^{*}_{ab}n^b$, being $n^a=(1,-\beta^i)/\alpha$ the normal to the spatial hypersurfaces. 
Notice the magnetic field differs in a sign respect to the usual definition, which was the one employed on \cite{FFE,FFE2}. 
The spatial Poynting vector is defined as,
\begin{equation}
 S^i := \epsilon^{ijk} E_j B_k   \label{S-def}
\end{equation}
where $\epsilon^{ijk}$ is the hypersurface induced volume element.

The evolution system is finally written as follows:
\begin{widetext}
\begin{eqnarray}
 \partial_{t} \phi &=& \beta^{i} \partial_{i} \phi + \alpha \mathcal{D}_j B^{j} - \alpha \kappa \phi  + \frac{\alpha}{\Delta^2} \tilde{E}^k r_k  \label{evol:phi}\\
 \partial_t E^{i} &=&  \left(\delta_{k}^{i} -\frac{\tilde{B}^{i}\tilde{B}_{k}}{\tilde{B}^2}\right) \left[\beta^{k} \mathcal{D}_j E^{j} + \mathcal{D}_j (\alpha F^{kj})  \right]   
 + \frac{\alpha \tilde{S}^{i}}{\tilde{B}^2} \mathcal{D}_j E^{j}  
 + \frac{\tilde{B}^{i}}{\tilde{B}^2} \left[ \beta^j r_{j} - \beta^k \tilde{E}_{k} \mathcal{D}_j B^{j} -\tilde{E}_{k} \mathcal{D}_j (\alpha F^{kj}) - \alpha \tilde{E}^j \partial_j \phi \right] \label{evol:E}\\
 \partial_{t} B^{i} &=&  \mathcal{D}_j (\alpha F^{*ij})  +  \beta^{i} \mathcal{D}_j B^{j}  + \alpha \gamma^{ij} \partial_{j} \phi 
                         - \frac{\alpha}{\Delta^2} \left[ \epsilon^{ijk} \, r_j \tilde{B}_k  + \frac{\tilde{E}^i}{\tilde{B}^2} \tilde{S}^k r_k \right] \label{evol:B} 
\end{eqnarray}
\end{widetext}
where we have denoted $\mathcal{D}_j (\cdot) \equiv \frac{1}{\sqrt{\gamma}} \partial_j (\sqrt{\gamma} \text{  } \cdot \text{  })$ and
\begin{eqnarray*}
&& r_i := \frac{\alpha^2}{4} \left[ \partial_i (G/\alpha^2) + \sigma \, \partial_i (F/\alpha^2) \right] 
\quad \text{;} \quad  \Delta^2 := \tilde{B}^2 - \tilde{E}^2  \\
&& \tilde{E}^i = E^i - \sigma B^i  \quad \text{;} \quad \tilde{B}^i = B^i + \sigma E^i \quad \text{;} \quad \tilde{S}^i = (1+\sigma^2 ) S^i 
\end{eqnarray*}

To have further control on the constraint $E\cdot B=0$, we adopt a damping strategy taken from \cite{alic2012},
\begin{equation}
 \partial_t E^i \rightarrow \partial_t E^i  - \alpha \, \delta \, \frac{E\cdot B}{B^2} B^i \label{damp}
\end{equation}
with a moderate coefficient $\delta \sim 100$ to enjoy the constraint cleaning properties, 
while avoiding the complications of having stiff terms which would demand implicit-explicit schemes (as pointed out in \cite{Pfeiffer2015}). 
In order to deal with current sheets we employ a ``standard'' approach in which electric field is effectively dissipated in order
to maintain the condition that the plasma is magnetically dominated, as discussed in \cite{FFE2}.

\subsection{Boundary Conditions}

This section is devoted to discuss the physical conditions at the global inner and outer boundaries of our domain and how to implement them numerically via the \textit{penalty method}. Additionally, we also incorporate an approach introduced in~\cite{FFE2} to restrict possible incoming violations of the divergence-free constraint, $\nabla \cdot B = 0$~\footnote{
We refer the interested reader to \cite{FFE2} (specifically, eqn's (3)-(4)) and also to ref.~\cite{Mari}, for more details on this method.}.

Our implementation of the penalties at the global boundaries is motivated by the interface treatment \eqref{interfaces}, 
identifying two main options: either one sets the incoming characteristic fields to a fixed source, regardless of the dynamics in the interior,
or one may use the information leaving the system by setting the incoming physical modes to a particular combination of outgoing ones.
The first choice has been already employed in \cite{FFE2} at the exterior surface, far away from the source, and we shall use it again here for the outer boundary. However, this approach is not suitable for setting the physical
conditions at the stellar surface. Instead, at the inner boundary we adopt the second approach, with a very specific combination of outgoing modes, as we will describe next.

\subsubsection{Inner boundary}

We shall think the stellar surface as a perfectly conducting layer that separates the force-free exterior from the interior of the star, 
and let us denote by $\mathcal{S}$ its three-dimensional world-volume. 
Our domain extends from the surface exterior $\mathcal{S}^+$, in the force-free
regime. However, the boundary conditions will be induced by the perfectly
conducting fluid at the surface interior $\mathcal{S}^-$.

Continuity across $\mathcal{S}$ of the Faraday equation, $dF = 0$, 
guarantees there are no jumps in the pullback to $\mathcal{S}$ of the electromagnetic tensor. Notice however that discontinuities on the remaining components of $F$ may result from the presence of surface charges and/or current densities at the conducting layer. The jump conditions $[A] \equiv A^+ - A^-$ at the interface can be written in terms of the co-moving electric and magnetic field $e_a := F_{ab} u^b$ and $b_a := F^{*}_{ab} u^b$ (i.e., where $u^a$ is the four-velocity of the co-rotating plasma), namely
\begin{eqnarray}
  {\bf n} \cdot [{\bf b}] &=& 0 ~~~,~~~ {\bf n} \cdot [{\bf e}] = q  \\
  {\bf n}  \times [{\bf b} ] &=& {\bf j} ~~~,~~~ {\bf n}  \times [{\bf e} ] = 0
  \label{conductor}
\end{eqnarray}

The definition of the fields at $\mathcal{S}^-$ arise from interior structure of the magnetic field in the star and the perfect conductor condition for the electric field $e_a = 0$. These values propagate for some of the components to the exterior solution through the junction conditions, namely
\begin{eqnarray}
  {\bf n} \cdot {\bf b}^+ &=& {\bf n} \cdot {\bf b}^- = f(r,\theta,\phi) ~~,~~ \\
  {\bf n}  \times {\bf e}^+ &=& {\bf n}  \times {\bf e}^- = 0
\end{eqnarray}
where $f(r,\theta,\phi)$ is an arbitrary function defined by
the magnetic field structure in the interior of the star.
In addition, as noticed in \cite{Gralla2014}, the force-free condition ${\bf e}\cdot {\bf b} =0$ implies the normal component of the co-rotating electric field must vanish as well 
(i.e.~ ${\bf n}  \cdot {\bf e}^+ = 0$, there are no induced charges) whenever the magnetic field is not tangential to the stellar surface. 


On the other hand, the ideal MHD condition $e_a=0$ also allows to easily translate the conditions applied to the co-moving EM fields to the fiducial ones, namely $E_i = - \epsilon_{ijk} v^j B^k$.
The velocity vector $v_i$ is constructed as the projection
of the four-velocity $u^a = W (n^a + v^a)$, being $W=(1-v_i v^i)^{-1/2}$ the Lorentz factor.

Let us assume then an axi-symmetric star, with a surface located at $r=R$ (i.e., such that the normal vector is radial), that rotates with four-velocity $u^a = k^a + \Omega \, \eta^a $, 
being $k^a \equiv (\partial_t )^a$ and $\eta^a \equiv (\partial_{\phi} )^a$ the two Killing vector fields of the Kerr spacetime. 
Thus, the resulting boundary conditions for the electric and the magnetic fields can be written as,
\begin{equation}
  B^r = f(r,\theta,\phi) ~~,~~
 \alpha \, E_{i} = \epsilon_{ijk} \left( \beta^j + \Omega \, \eta^j \right)  B^{k} \label{E-cond}
\end{equation}
where $E_r$ is to be enforced only if the magnetic field is not completely tangent to the star (i.e.~$B^r \neq 0$). Although written on a slightly different notation, the prescription for the tangential components of the electric field in \eqref{E-cond} is equivalent to the given at \cite{petri2015general}. 
Notice also that the remaining components of the magnetic field are free: the junction conditions involve unknown surface currents, so they can not be fixed.

To numerically implement these boundary conditions we proceed as follows.
We will keep the normal magnetic field fixed to its interior value, $B^r = B^{r}_o (\theta)$, 
by enforcing it at each Runge-Kutta substep as done in ref.~\cite{PHAEDRA}.
The electric field components \eqref{E-cond}, on the other hand, are going to be prescribed
by applying the \textit{penalty method} to the incoming physical modes. That is,
 \begin{equation}\label{inner-penalty}
  \partial_t U^{\mu} \rightarrow  \partial_t U^{\mu} + \frac{1}{ h \, \sigma_{o}} \sum_{a (\lambda_a > 0)} \lambda_{a} \, U^{\mu}_{(a)} L^{(a)}_{\nu} (U^{\nu}_o - U^{\nu})
 \end{equation}
where $L^{(a)}(\cdot)$ are a set of operators that must have a very precise structure in order to guarantee control of the semidiscrete energy by the penalties, namely:
\begin{equation}\label{boundary_op}
 L^{(a)}(\cdot) := \Theta^{a}_{>} (\cdot) - R^{a}_{~b} \Theta^{b}_{\leq} (\cdot)
\end{equation}
Here the co-basis elements $\Theta^{a}_{>}$ represents incoming modes, while $\Theta^{b}_{\leq}$ the outgoing and zero modes.
The idea is that one prescribes incoming physical modes at the boundary, combining information of the remaining characteristic modes.
The coefficients $R^{a}_{~b}$ might be interpreted as reflexion coefficients and must be solved according to the particular condition one wishes to impose.
A guidance in doing so, is provided by an observation regarding the action of the penalties on the boundary fields.
Namely, that they will dynamically enforce: 
\begin{equation}
 L^{(a)}_{\nu} (U^{\nu}_o - U^{\nu}) = 0
\end{equation}
from which it is deduced that the operators $L^{(a)}(\cdot)$ must project into the space of fields components to be specified
(in our case, for example, the electric field components involved in \eqref{E-cond}). 
While $U_o$ will act as a source, allowing to describe inhomogeneous boundary conditions.

In Appendix~\ref{sec:apxA}, we provide the reader with the specific boundary operators, $ L^{(a)}(\cdot)$, to enforce condition \eqref{E-cond}.
We construct these operators, going from the electro-vacuum case (i.e. Maxwell theory) to the full force-free system \eqref{evol:phi}-\eqref{evol:B}.
In between, we will discuss a simpler version of force-free electrodynamics often used, that might help to understand the transition to the full theory.

\subsubsection{Outer boundary}

The implementation of the outer boundary condition might be thought as a fictitious interface with an external field, $U_{ext}$. That is, setting the penalties as:
 \begin{equation}
  \partial_t U^{\mu} \rightarrow  \partial_t U^{\mu} + \frac{1}{ h \, \sigma_{o}} \sum_{a (\lambda_a > 0)} \lambda_{a} \, P^{\mu}_{(a)\nu} \left( U^{\nu}_{ext} - U^{\nu} \right)
 \end{equation}
which prescribe the incoming physical modes according to a fixed source we control. 
Thus, in ref.~\cite{FFE2} for instance, $U_{ext}$ represented a uniform magnetic field threatening the BH magnetosphere, sourced by a distant accretion disk.
As in the pulsar case there is no expected external electromagnetic sources, we will set $U_{ext}^{\nu} = 0$, corresponding to \textit{maximally dissipative} boundary conditions.
This means no physical signals would enter through the outer surface and all waves reaching it will leave the domain without reflections.
In practice, we will bring the exterior field smoothly to zero from its initial data configuration value. 
This way, we prevent the penalties from introducing spurious perturbations that would propagate inwards. 

\subsection{Initial Data}

We describe the spacetime around the rotating neutron star by the Kerr metric, which is parametrized by the mass $M$ and spin $a$. 
The metric in Kerr-Schild form can be written as,
\begin{equation}
 ds^2 = \left( \eta_{a b} + H \, \ell_{a } \ell_{b } \right)  dx^{a }dx^{b}
\end{equation}
where $\eta_{ab}$ is the flat metric and $\ell_{a}$ is a null co-vector with respect to the flat and the whole metric. 
In Cartesian coordinates $(t, x, y, z)$\footnote{Sometimes referred as the Kerr-Schild Cartesian coordinates or the Kerr-Schild frame (see e.g.~\cite{Visser2007}).}, the metric function $H$ takes the form, 
\begin{eqnarray}
H &=& \frac{2 M r}{r^2 + a^2 z^2 /r^2}\\
r^2 &=& \frac{1}{2}(\rho ^2-a^2) + \sqrt{\frac{1}{4}(\rho ^2-a^2)^2 + a^2 z^2} \\
\rho^2 &=& x^2+y^2+z^2  
\end{eqnarray}
and the co-vector $\ell_{a}$ reads
\begin{equation}
 \ell_{a} = \left\lbrace 1, \frac{rx + ay}{r^2+a^2}, \frac{ry-ax}{r^2+a^2}, \frac{z}{r} \right\rbrace .
\end{equation}



Assuming an homogeneous and spherically symmetric neutron star of radius $\rho=R$,
we can fix the dimensionless moment of inertia $\mathcal{I} := I/MR^2$ to the value $2/5$ 
(see e.g.~\cite{petri2015general, gralla2016pulsar}). This choice allow us to relate the star angular velocity $\Omega$ with the spin parameter in units of the stellar radius, namely 
\begin{equation}
 \frac{a}{R}=\frac{2}{5} \, \Omega \, R
\end{equation}
Notice that, for the spin range of realistic pulsars $a/R\lesssim0.15$, the Kerr spacetime is rather close to the metric of a neutron star in the slowly rotating limit, which is often used in the literature (e.g.~\cite{petri2015general, belyaev2016spatial,gralla2016pulsar, gralla2017inclined}).
Indeed, their metric components differ by less than $1\%$ from ours in this range. 

There are two other important dimensionless quantities that characterize the problem. The first one is the surface rotation velocity,
\begin{equation}
 v_s := \frac{R}{R_{LC}} 
\end{equation}
which ranges from $10^{-4}$ to $10^{-1}$ in realistic pulsars.
Here $R_{LC} = \Omega^{-1}$ denotes the usual light cylinder radius in flat spacetime.
The second quantity is the star compactness,
\begin{equation}
 \mathcal{C} := \frac{M}{R}
\end{equation}
which coincides with the definition in Ruiz et al. \cite{ruiz2014} for their compaction, 
and differs in a factor $2$ with other definitions like those in refs~\cite{petri2015general,gralla2017inclined,belyaev2016spatial}.
The compactness has a theoretical upper limit given by $\mathcal{C} < 4/9$ \cite{shapiro1983black}, and for typical model of neutron star interiors its value is around $\mathcal{C}\sim 0.2$.

We shall assume the magnetosphere is initially populated with a dipolar magnetic field given by the potential \cite{Shapiro1983},
\begin{equation}
 A_{\phi} = \frac{3\mu \sin^2 \theta}{4M} \left[ 1 + \frac{r^2}{2M^2} \ln (1-2M/r) +  \frac{r}{M} \right] \label{A-potential}
\end{equation}
where $\mu$ is the magnetic dipole moment. 
The electric and magnetic fields arising  in our foliation of Schwarzschild spacetime are just: 
\begin{eqnarray*}
 && B_{o}^{r} = \frac{3 \mu  \cos \theta \left(r^2 \log \left(1-\frac{2 M}{r}\right)+2 M (M+r)\right)}{4 M^3 r^2 \sqrt{\frac{2 M}{r}+1}} \label{Br}\\
 && B_{o}^{\theta} = \frac{3 \mu  \sin \theta  \left(2 M (M-r)+r (2 M-r) \log \left(1-\frac{2 M}{r}\right)\right)}{4 M^3 r^2 (r-2 M) \sqrt{\frac{2 M}{r}+1} } \label{Bth}\\
 && E_{o}^{\phi} = \frac{3 \mu \left(2 M (r-M)+r (r-2 M) \log \left(1-\frac{2 M}{r}\right)\right)}{2M^2 r^3  (r-2 M) \sqrt{\frac{2 M}{r}+1} } 
\end{eqnarray*}
whereas the other components vanish. Notice that this is an exact stationary solution of Maxwell vacuum equations for Schwarzschild but not for Kerr and that the magnetic field will only approximately  satisfy the constraint $\nabla \cdot B = 0$ on Kerr spacetimes. However, 
its value is dynamically kept very close to zero by using the divergence cleaning approach and constraint preserving boundary conditions. We will show evidence on such dynamic control of the constraint later on (see figure \ref{fig:constraint}). 

Notice also that in the Newtonian limit $\frac{r}{2M}\gg 1$ (where $\alpha \rightarrow1 $ and $\beta^i \rightarrow 0$) the magnetic field reduces to the usual flat space result in an orthonormal tetrad $\hat{e}_{\lambda}$,
\begin{eqnarray}
  B_{o}^{\hat{r}} = \frac{2 \mu  \cos \theta }{r^3} ~~,~~
  B_{o}^{\hat{\theta}} = \frac{\mu  \sin \theta }{r^3} 
\end{eqnarray}
with vanishing electric field. 

In order to avoid transient initial currents and sharp profiles of the fields, the star will be smoothly brought from rest to its final angular velocity $\Omega$ by using a time-dependent function,
\begin{equation}
  \frac{1}{2} \Omega \left[ 1 - \cos (\pi \, t/t_s ) \right] ~~~~if~~~ 0 \leq t \leq t_s        
\end{equation}
with $t_s$ usually chosen to be around $1/5$ of the rotation period.  
To abandon axial symmetry we shall simply tilt the magnetic dipole axis of the initial configuration on the $x-z$ plane, 
while keeping the rotation axis of the star (i.e., and the spacetime) unchanged on the $z$-direction. 
Tilting the dipole axis corresponds to the following change of Cartesian coordinates,
\begin{eqnarray*}
 && x' = x \cos \chi  + z \sin \chi\\
 && y' = y \\
 && z' = z \cos \chi  - x \sin \chi  
\end{eqnarray*}
being $\chi$ the inclination angle. Whereas for imposing the normal component of the magnetic field at the inner boundary, it will be necessary to keep track of the accumulated rotation phase during the evolution, and apply it to the initial field to get the boundary data for $B^r$.

\subsection{Analysis Quantities}

We shall compute the spin-down luminosity as the integrated Poynting flux on spherical shells,
\begin{equation}
 L := \int p^r \, \sqrt{-g} \, d^2 w
\end{equation}
where $p^a := -T^{ab}k_b $ is the conserved electromagnetic four-momentum
(being $T^{ab}$ the electromagnetic stress-energy tensor and $k^a = (\partial_t )^a$ the Killing vector field), $p^r \equiv p^a (dr)_a $ its radial component 
and $d^2 w$, the differentials of the angular coordinates (see e.g.~\cite{Palenzuela2010Mag,FFE2}) \footnote{Notice 
this definition is consistent with the one in \cite{petri2015general} written on an slightly different language, namely:  $L:= \int_{\mathcal{S}_L} (E \wedge H)^{\hat{r}} \, dS $.}.
We will consider the spin-down luminosity to be the one computed at the light cylinder, as in Petri \cite{petri2015general}; 
while Ruiz et al.~\cite{ruiz2014} has taken instead the asymptotic value, which might depend on the numerical resistivity and dissipation of the numerical scheme at the current sheet. 

The spin-down power of the aligned rotator has been estimated analytically in flat spacetime, yielding a dependence proportional to $\mu^2 \, \Omega^4$ (in geometrized units). This relation turned-out to be quite accurate, agreeing within a factor close to unity with most numerical simulations in flat spacetime (see e.g.~\cite{cerutti2017} and references therein). Therefore, it is common to set
\begin{equation}\label{Lo}
 L_o = \mu^2 \, \Omega^4
\end{equation}
as a reference luminosity in order to rescale the results.
The power of $\Omega$ appearing on \eqref{Lo} is directly related with another interesting quantity, known as the braking index $n$,
defined as $\dot{\Omega} = -K \, \Omega^{n}$, or, equivalently, $n=\Omega \ddot{\Omega}/\dot{\Omega}^2$.

Since the spin-down luminosity is associated to the star rotational energy loss by $L =-I \, \Omega \,\dot{\Omega}$, one can formally derive $n$ in terms of the luminosity and the surface rotation velocity $v_s$, as done in ref.~\cite{petri2015multipolar},
\begin{equation}\label{n}
 n = \frac{\Omega}{L} \frac{dL}{d\Omega} - 1 =\frac{v_s}{L} \frac{dL}{dv_s} - 1
\end{equation}
The braking index measures how the spin period and its derivatives change with time,
and it can be inferred from the physical mechanism regulating the spin-down.
For the aligned point-dipole magnetic field, this quantity is known to give exactly
$n= 3$.  However, considering neutron stars of finite size --as well as allowing curvature effects  and  misalignment-- induces  deviations  from  this
value, as we shall see on Section~\ref{sec:break}.





\section{Numerical Results}\label{sec:results}

In this section we present a detailed numerical study of the pulsar magnetosphere, mainly focusing on how the general relativistic effects impact on the resulting spin-down luminosity. 
First we study the aligned case, when the magnetic dipole coincides with the spin of the star and the solution is axially symmetric. Our code reproduces the well known properties of the aligned rotator magnetosphere on flat spacetime. Basic tests of our numerical implementation are included here, as well as a convergence analysis of the luminosity for different numerical resolutions and evidences of the correct behavior of the solenoidal constraint. Later, we incorporate the full general relativistic effects and compare our results with previous studies.
We have carefully investigate the Poynting luminosity dependence on stellar surface velocity $v_s$ and compactness $\mathcal{C}$ (see \eqref{aligned-fit} and Fig.~\ref{fig:compactness});
and then generalize it to the case where the magnetic dipole and rotation axes are not aligned, thus adding the dependence with the misalignment angle $\chi$ (see \eqref{misaligned-fit} and Fig.~\ref{fig:inclination}).
Our main result is summarized in a fitted formula which is a good approximation for any star, depending only on three adimensional parameters: 
the surface velocity $v_s=R/R_{LC}$, the stellar compactness $\mathcal{C}=M/R$ and the misalignment angle $\chi$ between the spin and the magnetic dipolar moment.
Finally, we have used this generic formula to estimate corrections to the braking index in terms of these three parameters.

\subsection{Aligned Rotator}

\subsubsection{Magnetic field topology and tests}

\begin{figure}
  \begin{center}
\includegraphics[scale=0.25]{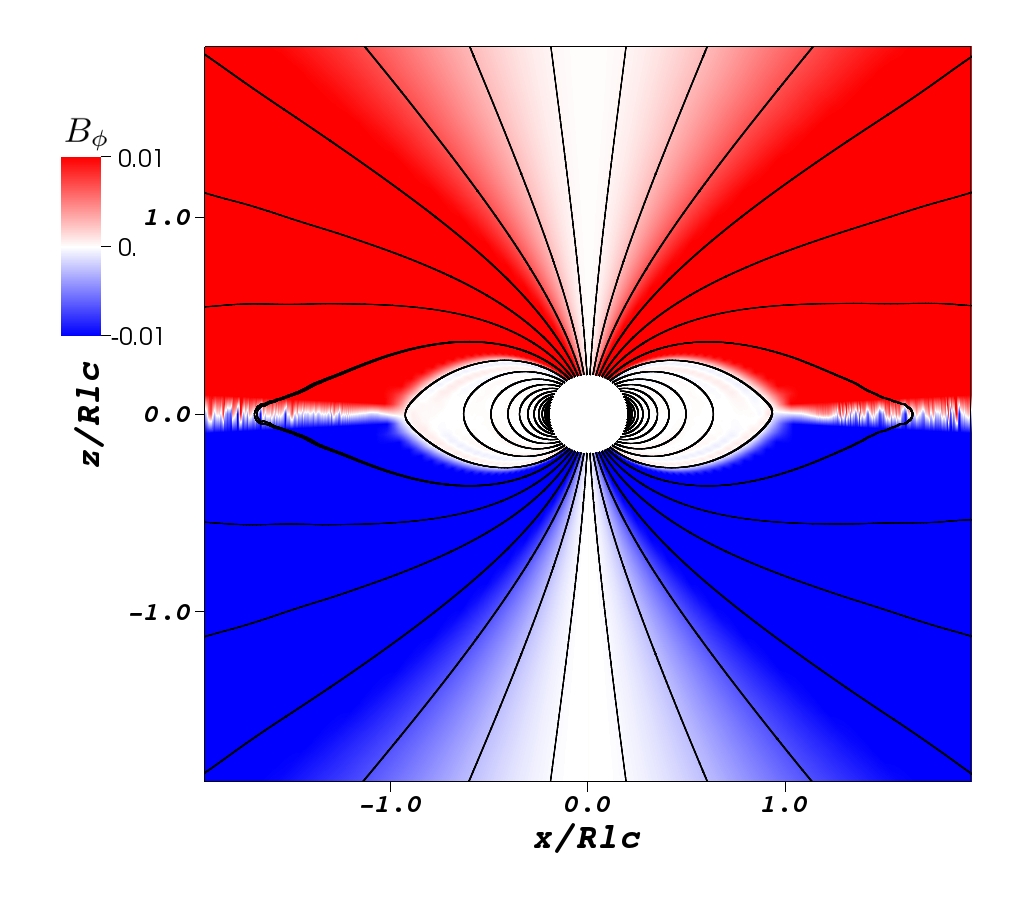}
  \caption{{\em Aligned Rotator}. Global structure of the aligned pulsar magnetosphere. 
  Black lines represents the poloidal magnetic field, while the color scale corresponds to the toroidal magnetic component, $B_{\phi}$.
  }
 \label{fig:aligned} 
 \end{center}
\end{figure}

Our simulations reproduce all the well known features of an aligned rotator in flat spacetime 
(see e.g.~\cite{komissarov2006simulations, mckinney2006relativistic, timokhin2006force, contopoulos2006, li2012, PHAEDRA}).
The late time solutions exhibit a closed zone that co-rotates with the star, extending up to the light cylinder.
This zone ends on a Y-point at the equatorial plane from where a strong equatorial current sheet begins. 
Our maximally dissipative boundary conditions allow us to place the outer numerical boundary fairly close to the central region, at around $4R_{LC}$.
The evolution reaches steady state after $\sim2$ rotational periods of the star. A typical equilibrium solution is depicted in Fig.~\ref{fig:aligned}, showing the global magnetic field structure of the magnetosphere: 
lines represents the poloidal magnetic field and color illustrates the toroidal component. 
Notice that some of the field lines close outside the light cylinder due to effective resistivity of the current sheet,
but the co-rotating region with vanishing toroidal magnetic field remains within the light cylinder.

The luminosity is constant between the stellar surface and the light cylinder, as expected from conservation of the electromagnetic energy-momentum tensor.
Beyond the light cylinder the energy flux slowly decreases with radius due to dissipation at the equatorial current sheet.
\begin{figure}[h!]
  \begin{center}
\includegraphics[scale=0.35]{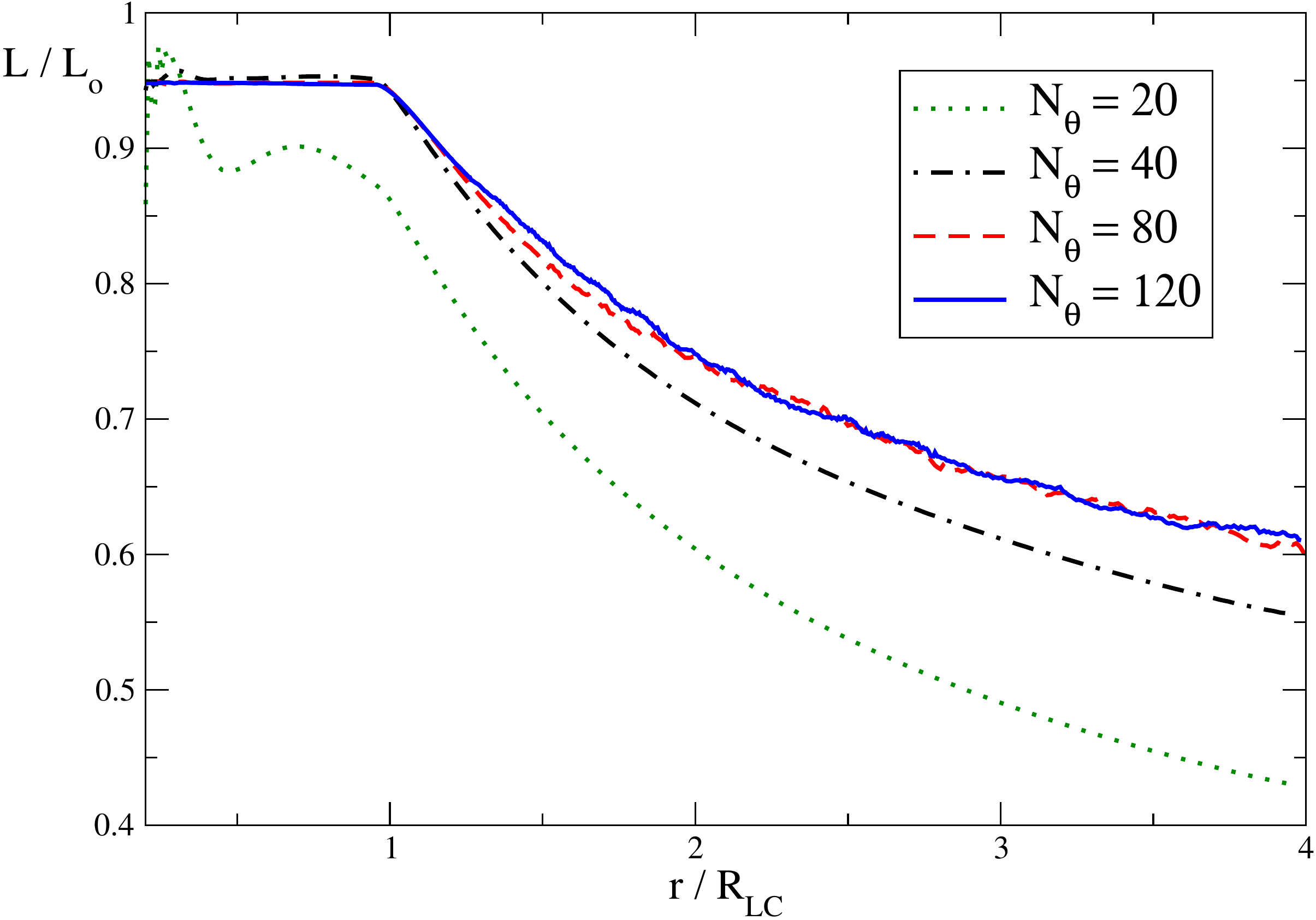}
  \caption{{\em Aligned Rotator}. Poynting luminosity as a function of radius,  computed with different numerical resolutions $N_{\theta} \times N_{\phi} \times N_{r}$ and using $N_{r} = 2 N_{\phi} = 4 N_{\theta}$.}
 \label{fig:resolution} 
 \end{center}
\end{figure}
In Fig.~\ref{fig:resolution} it is displayed the luminosity, integrated on spherical surfaces, as a function of the radius for a typical late-time solution of the aligned rotator. The curves corresponds to various numerical resolutions $N_{\theta} \times N_{\phi} \times N_{r}$ with $N_{r} = 2 N_{\phi} = 4 N_{\theta}$. Clearly, 
the flux is already converging for a moderate resolution $N_{\theta} = 40$.
We generally employ a higher resolution, with $N_{\theta} = 80$, for the results presented in this paper.
We confirmed numerically that the resulting luminosities, normalized conveniently with $L_o = \mu^2 \Omega^4$, do not depend on the given angular velocity $\Omega$ but only on the adimensional relation $v_s = R/R_{LC} \equiv R\, \Omega$.

\begin{figure}
  \begin{center}
\includegraphics[scale=0.35]{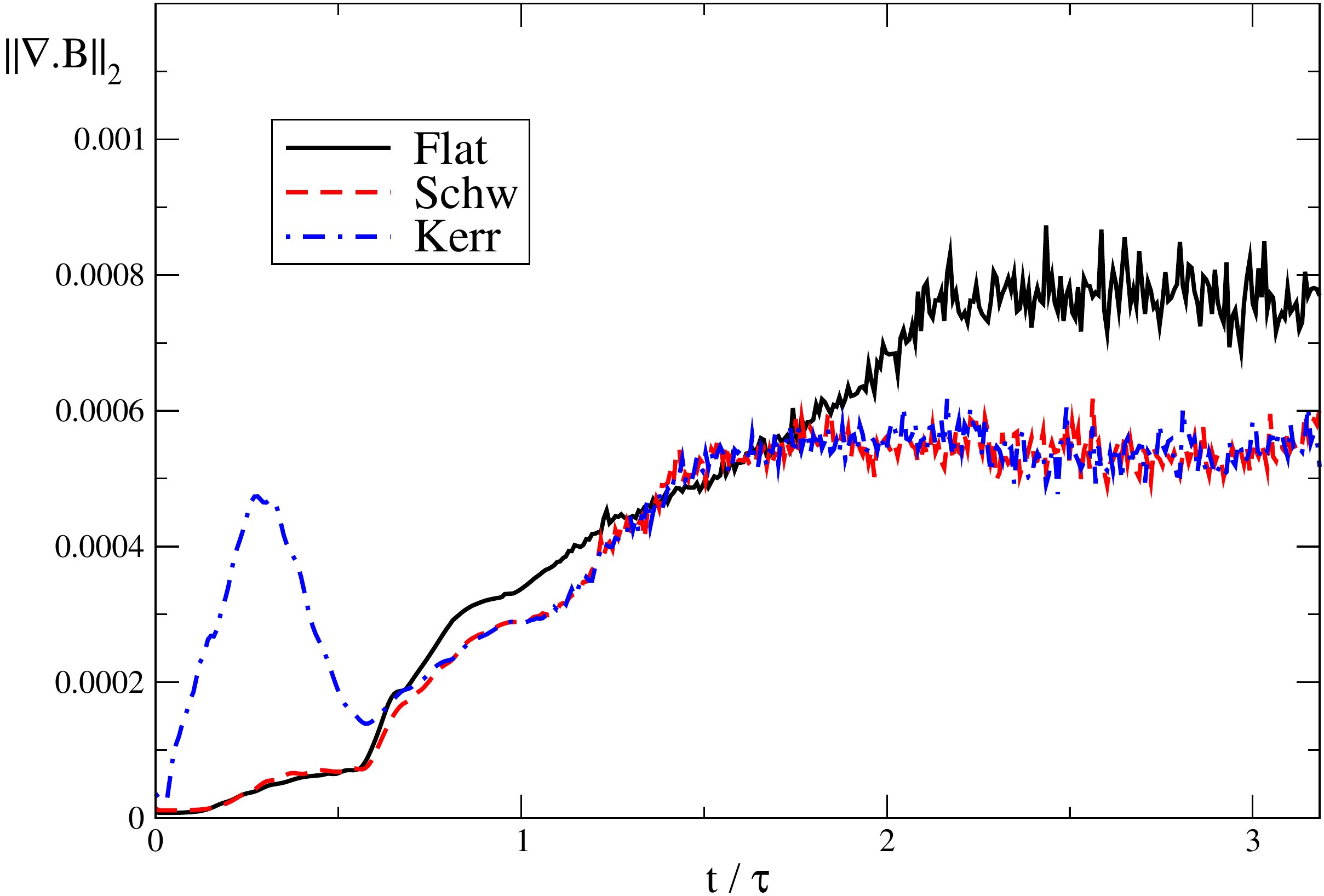}
  \caption{{\em Aligned Rotator}. The normalized divergence of the magnetic field $||\nabla \cdot B ||_2 / || B ||_2$ is plotted as a function of time (in units of the rotation period $\tau = 2\pi / \Omega$). The solenoidal constraint is kept under control by the divergence cleaning techniques.}
 \label{fig:constraint} 
 \end{center}
\end{figure}

The behavior of the solenoidal constraint $\nabla \cdot \vec{B} = 0$ on a typical run is displayed in Fig.~\ref{fig:constraint}, where the quotient $||\nabla \cdot B ||_2 / || B ||_2$ is plotted as a function of time for Kerr, Schwarzschild and flat spacetimes. As it can be seen, its value remains always smaller than $10^{-3}$, thus showing a decent control of the constraint during the evolution.
Even on Kerr spacetime --where the initial data slightly violates the solenoidal constraint-- we observe the divergence cleaning mechanism
act to keep its value small, and moreover, it ends-up driving it to a similar value to that of Schwarzschild after an initial transient.
\begin{figure}
  \begin{center}
\includegraphics[scale=0.25]{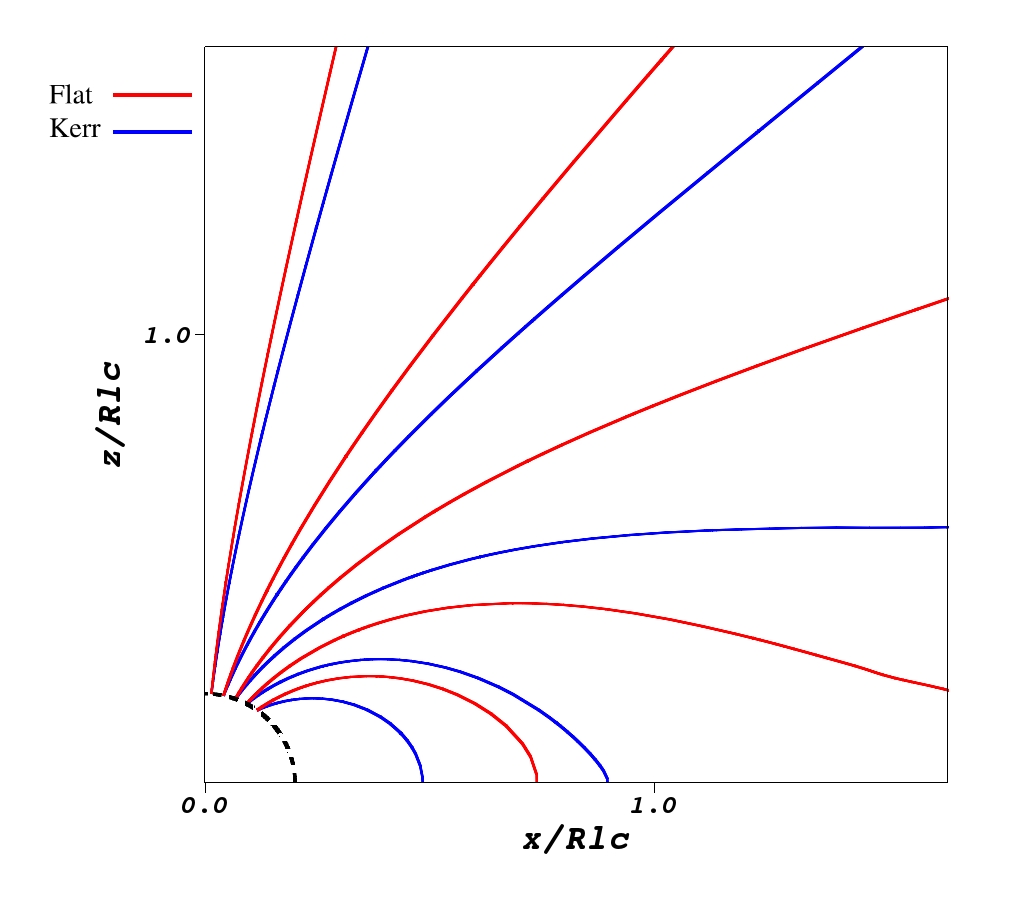}
  \caption{{\em Aligned Rotator}. Some representative poloidal field lines that illustrate the qualitative effects due to curvature. 
  Red lines depict Newtonian solutions, while blue ones correspond to the GR results.
  }
 \label{fig:aligned_comp} 
 \end{center}
\end{figure}

\subsubsection{Comparison with previous results}

For concreteness, we shall consider here full GR simulations at a high but still realistic star compactness $\mathcal{C}=0.25$. 
As previously observed in Ref.~\cite{petri2015general}, we find that including curvature effects does not significantly change the field topology. 
It does, however, compress the poloidal magnetic field lines towards the neutron star equator, as it can be seen in Fig.~\ref{fig:aligned_comp}. 
Gravitational effects have been also shown to considerably affect the polar cap structure (see e.g. \cite{philippov2015ab, belyaev2016spatial, gralla2016pulsar}). 
As a result, there is an enhancement of the spin-down luminosity, as we shall see below.
Figure \ref{fig:comparisons} displays the Poynting luminosities for several spin rates $v_s =\left\lbrace 0.025 \text{, } 0.05 \text{, } 0.1 \text{, } 0.2 \text{, } 0.3 \text{, } 0.4 \right\rbrace$
and compares with the results of Ruiz et al.~\cite{ruiz2014} and Petri \cite{petri2015general}. 
It can be seen that our numerical data is in very good agreement with these previous studies, despite using different formulation of force-free electrodynamics and numerical methods.
In particular, we notice that in the region comprised between $v_s \simeq 0.1$ and $v_s \simeq 0.3$ the three models coincides almost perfectly.
For this particular case with $\mathcal{C}=0.25$, we find an enhancement of the total emitted power due to relativistic effects of about $15$\% for $v_s =0.1$, $30$\% for $v_s =0.2$, and even larger for higher spin rates. 
It it worth mentioning that Petri reported a discrepancy with Ruiz et al. respect to the aforementioned increments. We do not observe such discrepancies and thus believe that it may be related with a misinterpretation of Ruiz numerical data, which used a rather different setting and notation.
%
\begin{figure}
  \begin{center}
\includegraphics[scale=0.35]{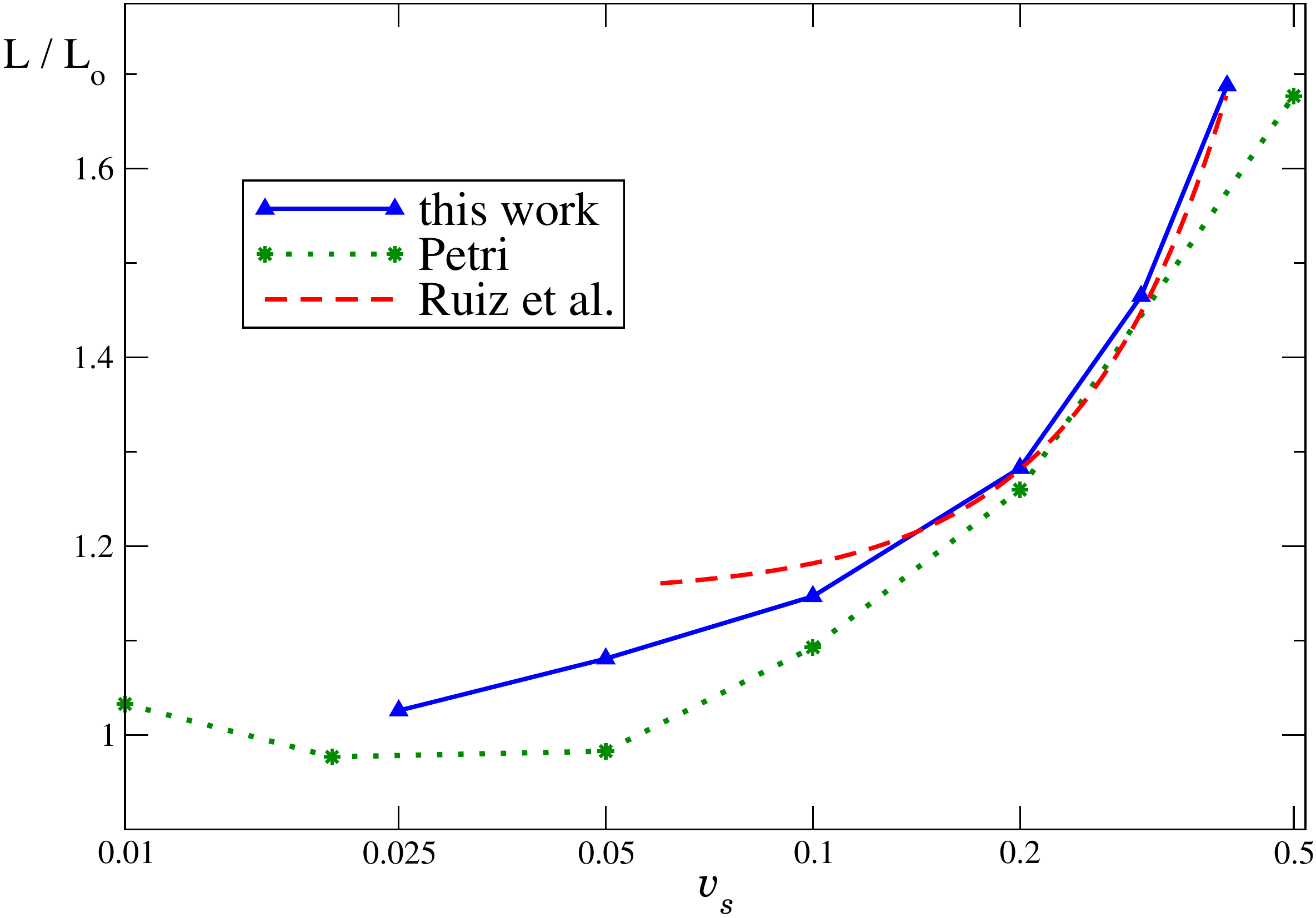}
  \caption{{\em Aligned Rotator}. Poynting luminosities (normalized with $L_o=\mu^2 \Omega^4$) that include the effects of general relativity at a fixed compactness, $\mathcal{C}=0.25$.
  Our results (blue triangles/solid-lines) are compared with previous numerical studies at several rotation rates $v_s$.
  Green asterisk/dotted-lines represents the data taken from Figure 22 of Petri \cite{petri2015general}, while that the red dashed curve was constructed by fitting the values from Table II of Ruiz et al.~\cite{ruiz2014} (evaluating at $\mathcal{C}=0.25$).
  }
 \label{fig:comparisons} 
 \end{center}
\end{figure}

\begin{figure}
  \begin{center}
\includegraphics[scale=0.35]{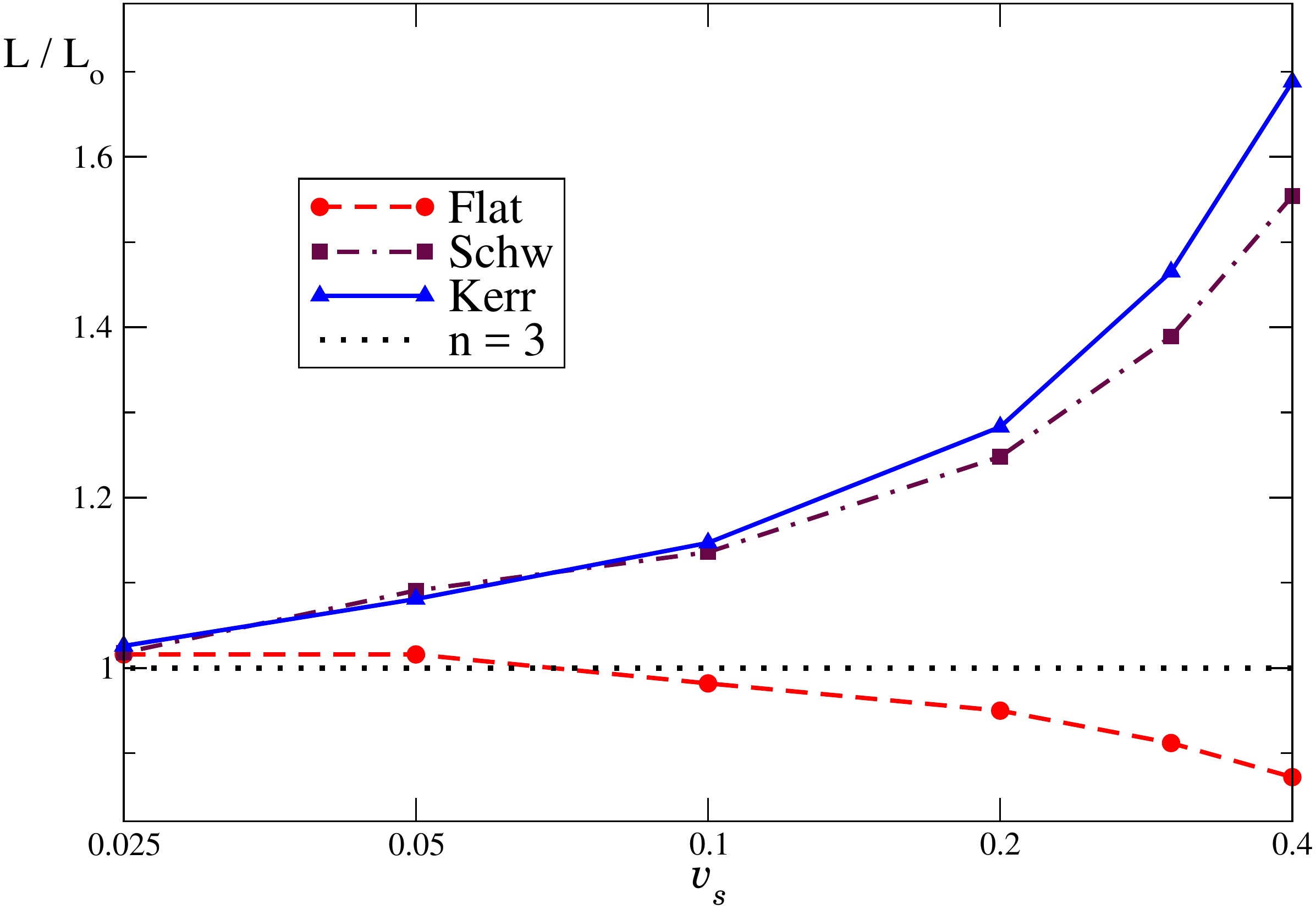}
\includegraphics[scale=0.35]{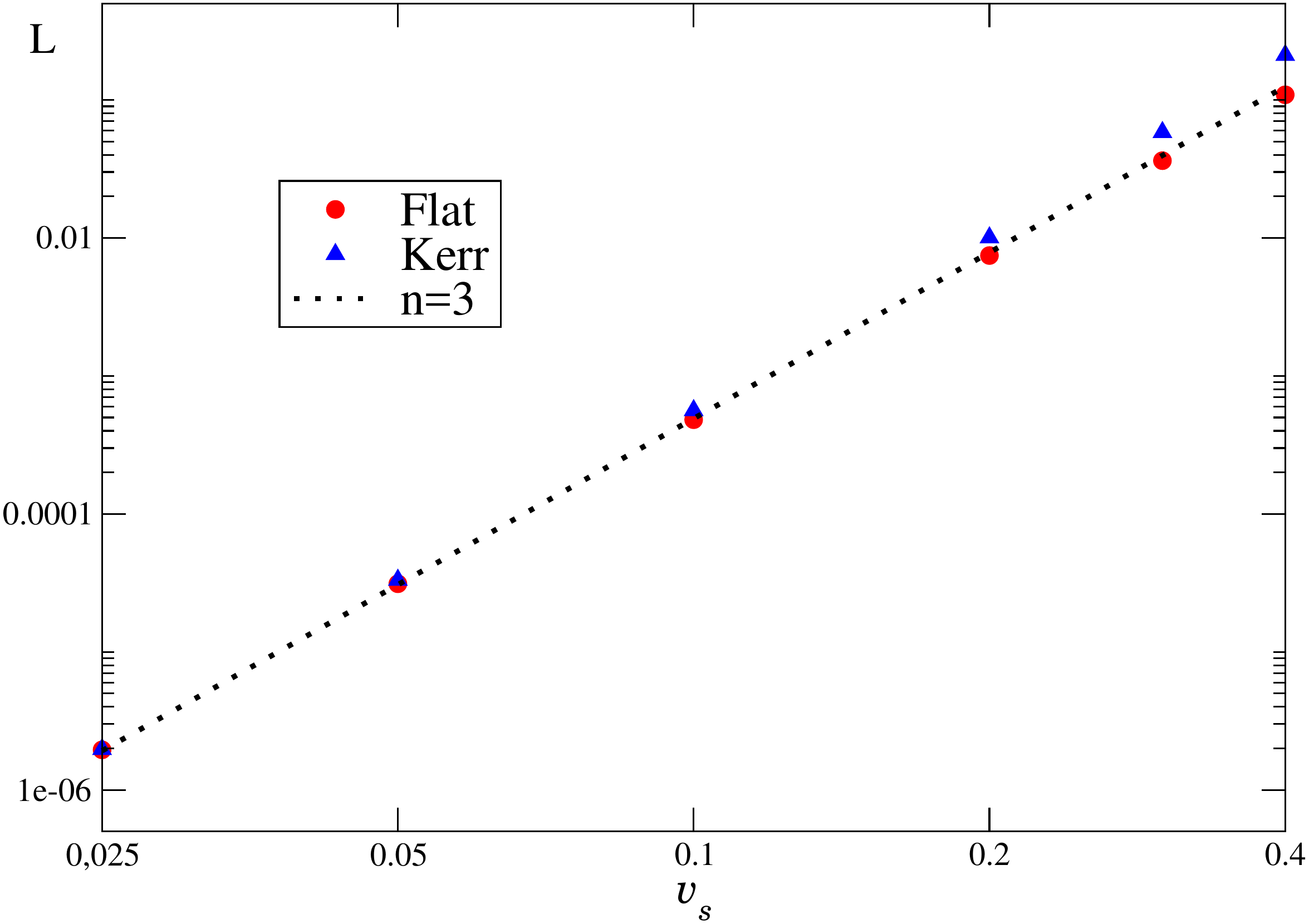}
  \caption{{\em Aligned Rotator}. Spin-down luminosity dependence on spin rate, $v_s$. 
  (Top panel) Comparison among Newtonian and GR (Schwarzschild and Kerr, at $\mathcal{C}=0.25$) solutions. 
  (Bottom panel) Effective braking indexes yield $n=2.94$ and $n=3.17$ for flat and Kerr spacetimes, respectively.}
 \label{fig:aligned_rotation}  
 \end{center}
\end{figure}

Figure \ref{fig:aligned_rotation} compares the luminosities resulting from full GR simulations with those on the Newtonian regime.
In the slow rotation limit, both GR and flat spacetime solutions seems to approach a similar value close to $L_o$.
This is consistent with the fact that relativistic corrections tends to disappear for small values of $v_s$, 
as pointed out recently by ref.~\cite{gralla2016pulsar} for comparisons made at fixed $\mu$ in the regime $R/R_{LC}\ll1$.
For spin rates $v_s \gtrsim 0.05$ deviations become apparent: Newtonian luminosities slightly decreases with rotation, while GR solutions increase their values significantly. 
Although one may distinguish among curvature effects associated with the spacetime mass and spin, they might be difficult to disentangle. 
We try to isolate the influence of stellar compactness from frame dragging effects by setting $a=0$ (i.e.~Schwarzschild metric).
The results are shown in the top panel of Fig.~\ref{fig:aligned_rotation}, where it can be noticed that frame dragging only produces a modest enhancement for $v_s\gtrsim 0.1$. 
It has been argued that the frame dragging effect is being compensated by an effective reduction of the angular velocity, which happens at the Kerr metric (see e.g.~\cite{philippov2015ab}) but not in Schwarzschild.
Bottom panel of Fig.~\ref{fig:aligned_rotation} presents the non-normalized luminosity as a function of the surface velocity on logarithmic scale, 
from where an effective braking index might be obtained. 
Our results compares well with those found by Petri \cite{petri2015general}, although we get slightly different values for the braking indexes:
we estimated $n=2.94$ for flat spacetime and $n=3.17$ for GR (Fig.~\ref{fig:aligned_rotation}, bottom image), whereas he gets $n=2.97$ and $n=3.12$, respectively~\footnote{A plausible explanation for this small discrepancy is that Petri has considered smaller rotation rates (like $v_s=0.01$) where the trend is very close to $n=3$. 
Thus, when fitting the curves he gets effective braking indexes closer to this value.}.

\subsubsection{Dependence on compactness and surface velocity}

\begin{figure}
  \begin{center}
\includegraphics[scale=0.35]{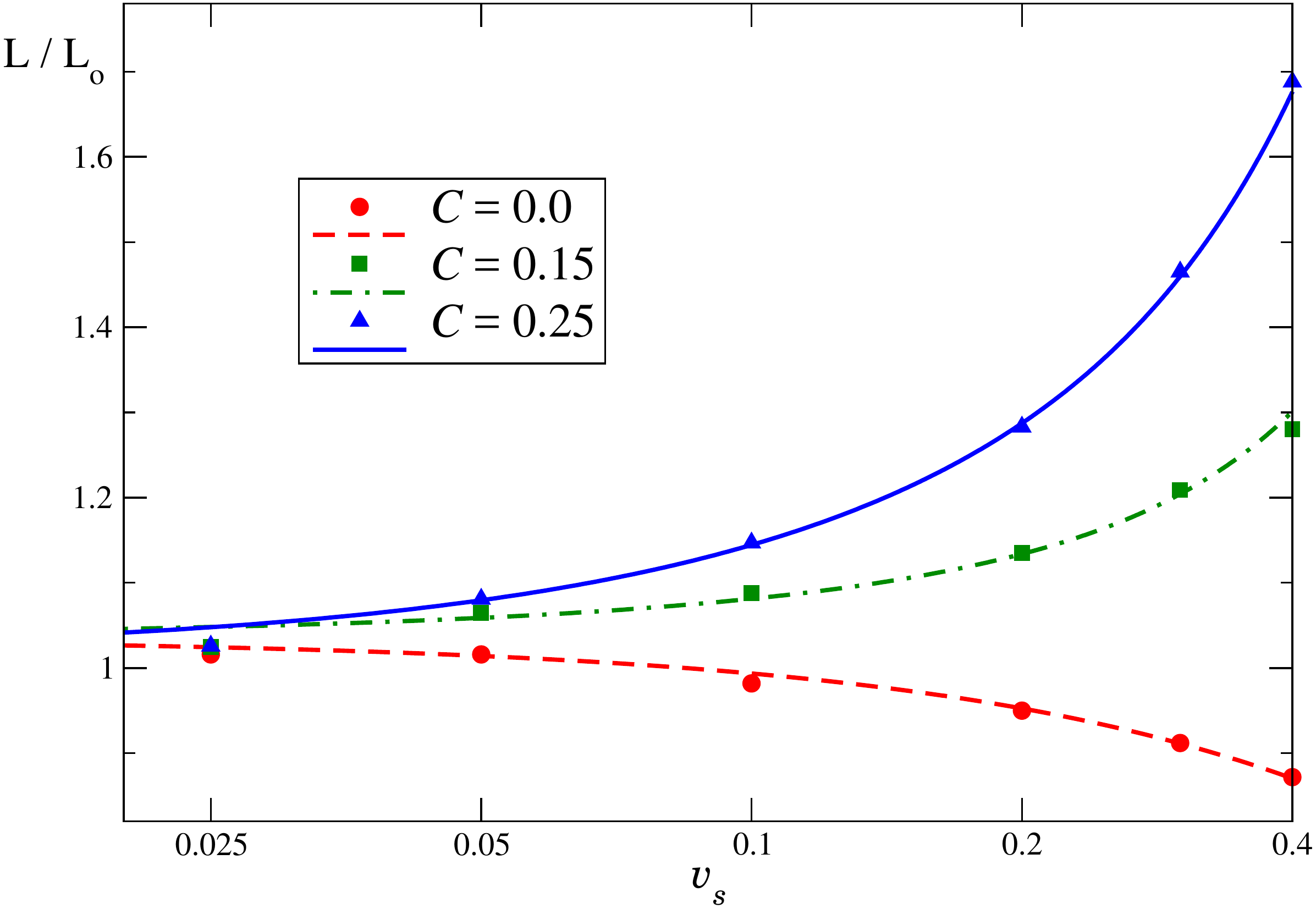}
\includegraphics[scale=0.35]{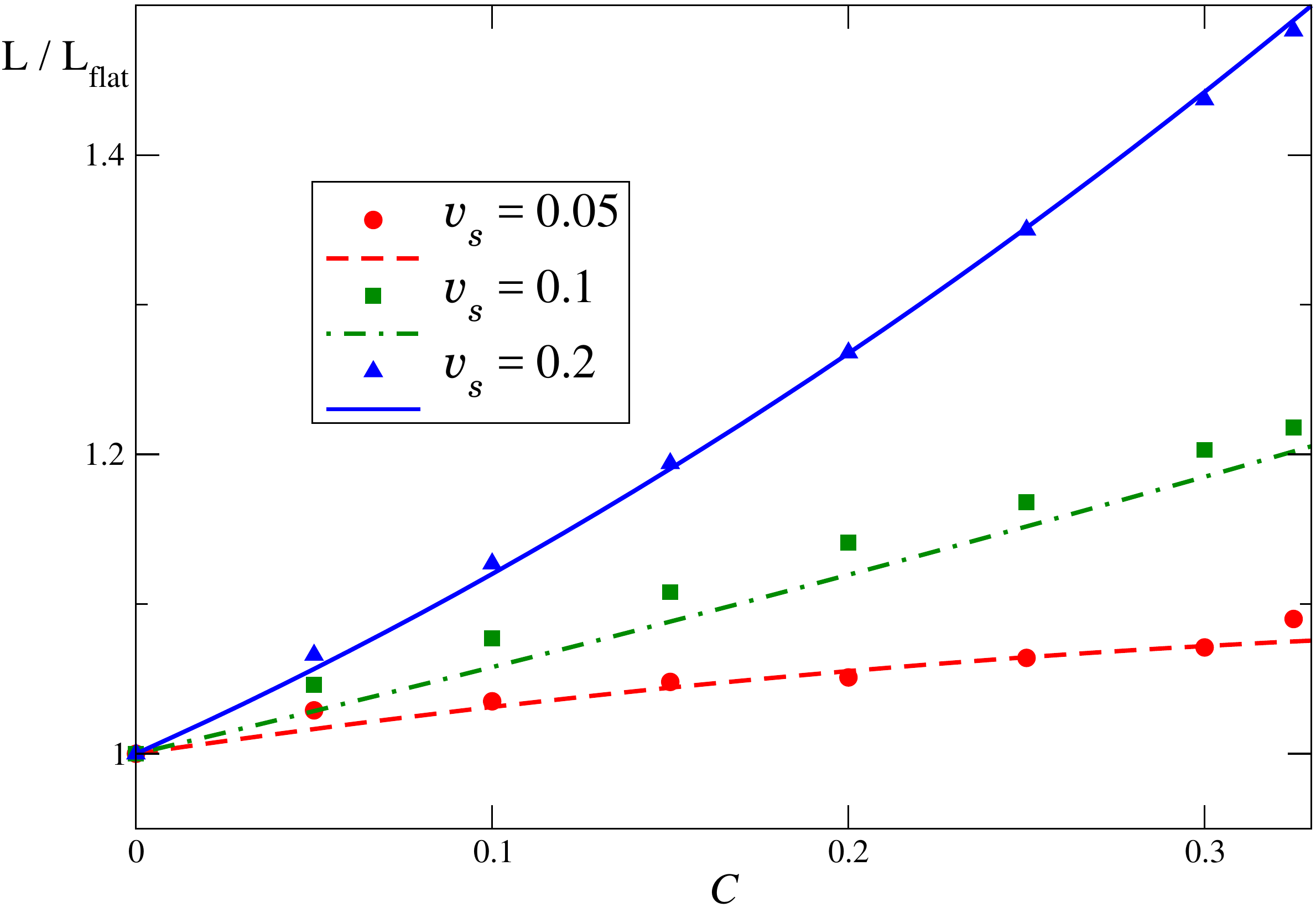}
  \caption{{\em Aligned Rotator}. Spin-down luminosity dependence on the surface velocity at $\mathcal{C} = \left\lbrace 0.0, 0.15, 0.25 \right\rbrace $ (\textbf{top image});
  and on stellar compactness for $ v_s = \left\lbrace 0.05, 0.1, 0.2 \right\rbrace $ (\textbf{bottom image}). 
  Numerical data has been displayed along with the relevant sections of the fit \eqref{aligned-fit}; and luminosities were normalized with $L_o=\mu^2 \Omega^4$ 
  and $L_{flat}\equiv L(\mathcal{C}\text{=0}, v_s )$, respectively.
  }
 \label{fig:compactness} 
 \end{center}
\end{figure}

In order to analyze quantitatively the general dependence of the spin-down luminosity on gravitational effects, we have considered different stellar compactness at various rotation rates. By using our numerical results,
we have been able to derive a fitting formula for the spin-down luminosity of an aligned rotator as a function of star compactness $\mathcal{C}=M/R$
and surface velocity $v_s = R/R_{LC}$.
An excellent fit of the numerical data was achieved with
\begin{eqnarray}\label{aligned-fit}
L (\mathcal{C}, v_s ) = L_o && \left\lbrace    1.035 - 0.41  \, v_s + (0.16 + 3.9 \, v_s) \, \mathcal{C} \right.  \nonumber\\
  && + \left. 9.8 \, v_{s}^3 \, \mathcal{C} + (11 \, v_s - 0.9) \, \mathcal{C}^2   \right\rbrace 
\end{eqnarray}
In Figure \ref{fig:compactness} there are displayed constant-$\mathcal{C}$ (top panel) and constant-$v_s$ (bottom panel) sections of the fit \eqref{aligned-fit},
along with their corresponding numerical data. These plots give an idea on how the aligned rotator luminosities vary on the two parameters $v_s$ and $\mathcal{C}$, as well as to illustrate the quality of the fitting formula obtained.
Clearly, the higher the compactness the larger the enhancement of the luminosity, with more impact on those stars that rotates faster.
Notice that the luminosity in the plot at the bottom panel has been normalized with the value for zero compactness (i.e., $L_{flat} \equiv L(\mathcal{C}\text{=0}, v_s ) = L_o (1.035 - 0.41  \, v_s $) for representation proposes only.
Finally, notice that we could find stable solutions for star's compactness up to $\mathcal{C}\approx 1/3$. Beyond that limit, simulations become unstable near the stellar surface, probably due to the presence of a light ring. Note however that this value is outside the range of allowed compactness ratios, as constrained by observations ($M\sim 1.1-2 M_{\odot}$, $R\sim 10-14$ km, so that $\mathcal{C}\sim 0.1-0.3$).

\subsection{Misaligned Rotator}

\begin{figure}
  \begin{center}
\includegraphics[scale=0.25]{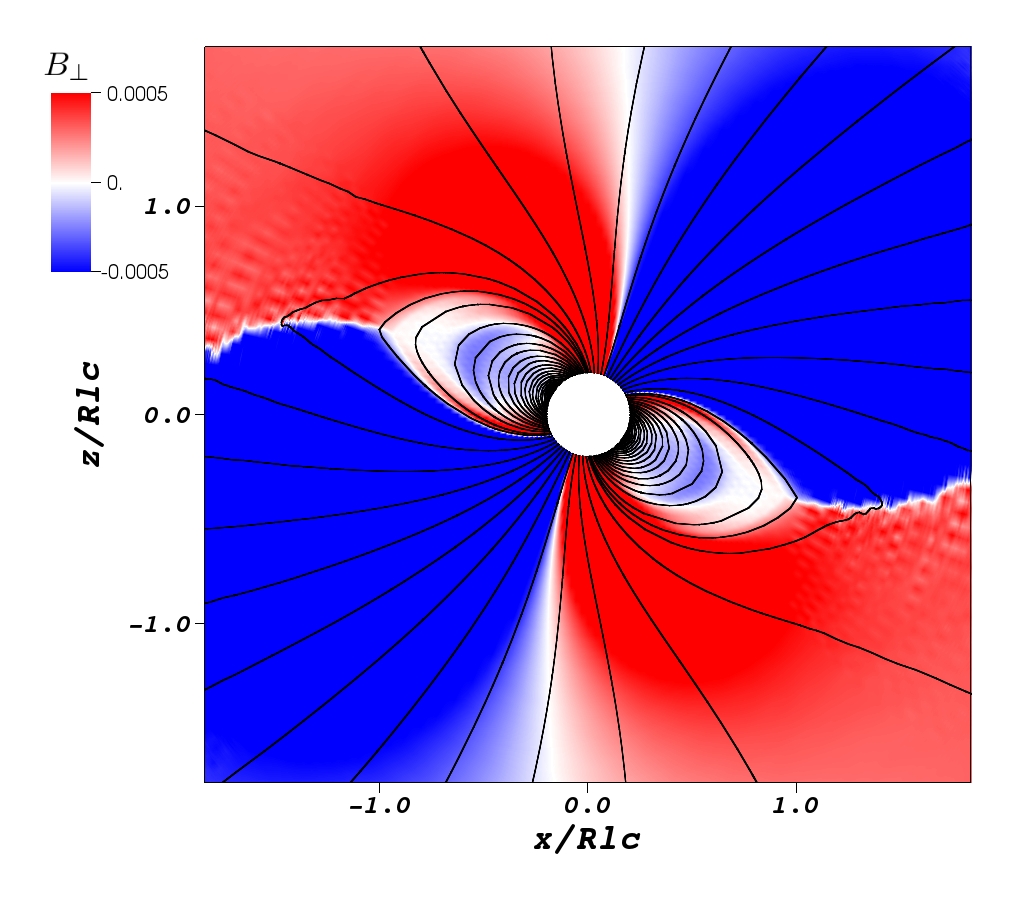}
  \caption{{\em Misaligned Rotator}. Poloidal and toroidal magnetic field obtained with $v_s = 0.2$ and misalignment angle $\chi=30$\textdegree.
  Lines represents the magnetic field in the ${\bf \mu - \Omega}$ plane, whereas the color scale corresponds to the magnetic component perpendicular to the plane.}
 \label{fig:misaligned} 
 \end{center}
\end{figure}

We study now the pulsar magnetospheres for the misaligned cases, occurring when the dipolar magnetic field of the star is inclined an angle $\chi$ with respect to its rotation axis. Again, our simulations reproduce previous results in the literature of a misaligned rotator in flat spacetime. Our numerical evolutions settle to steady state after about two stellar rotation periods. As it can be seen in Fig.~\ref{fig:misaligned}, the magnetic field topology on the ${\bf \mu - \Omega}$ plane is reminiscent of the aligned solution
with closed and open zones. The current sheet starts at the intersection of the closed zone with the light cylinder and oscillates around the rotational equator. The Poynting fluxes are again constant between the stellar surface and the light cylinder, where they are measured. The energy dissipation at the current sheet decreases as the misalignment angle increases, as reported in reference \cite{cao2016} for the Newtonian case.

\begin{figure}
  \begin{center}
\includegraphics[scale=0.25]{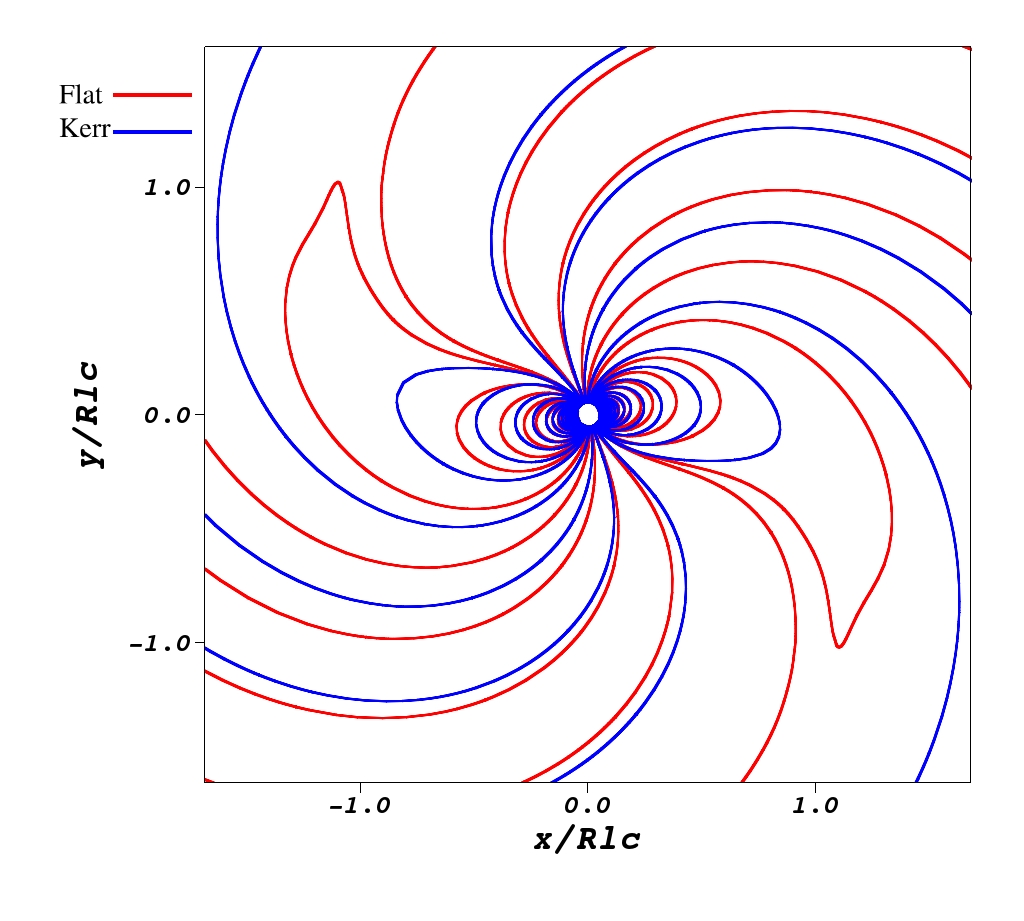}
  \caption{{\em Misaligned Rotator}. Equatorial field lines of the orthogonal rotator with $v_s = 0.05$.
  Red and blue lines corresponds to flat and Kerr spacetime solutions, respectively. }
 \label{fig:orthogonal} 
 \end{center}
\end{figure}

Including the effects of general relativity does not change dramatically this qualitative picture. 
In Figure \ref{fig:orthogonal}, the equilibrium solution of the orthogonal rotator in GR is compared to the one obtained in the Newtonian regime.
The equatorial field lines in both cases resemble the Deutsch solution inside the light cylinder and exhibit an spiral structure outside, 
being the GR (blue) lines more curved than the Newtonian (red) ones due to the   gravity pull. 
Quantitatively, the presence of spacetime curvature enhances the spin-down luminosity in the misaligned case, although it depends softly on the compactness and the surface velocity. We have evolved the magnetospheres varying the inclination angles $\chi$ for three different rotation rates, 
comparing the luminosities between the Newtonian and the general relativistic regime of a star with compactness $\mathcal{C}=0.25$.
The values of the luminosity obtained from our simulations were fitted by the following expression,
\begin{equation}\label{misaligned-fit}
\frac{L (\mathcal{C}, v_{s}, \chi)}{L(\mathcal{C}, v_{s}, \chi\text{=0})} = 1 + \left( 1.24 + 8.14 \, v_s \, \mathcal{C}\right)  \sin^2 \chi   
\end{equation}

\begin{figure}
  \begin{center}
\includegraphics[scale=0.35]{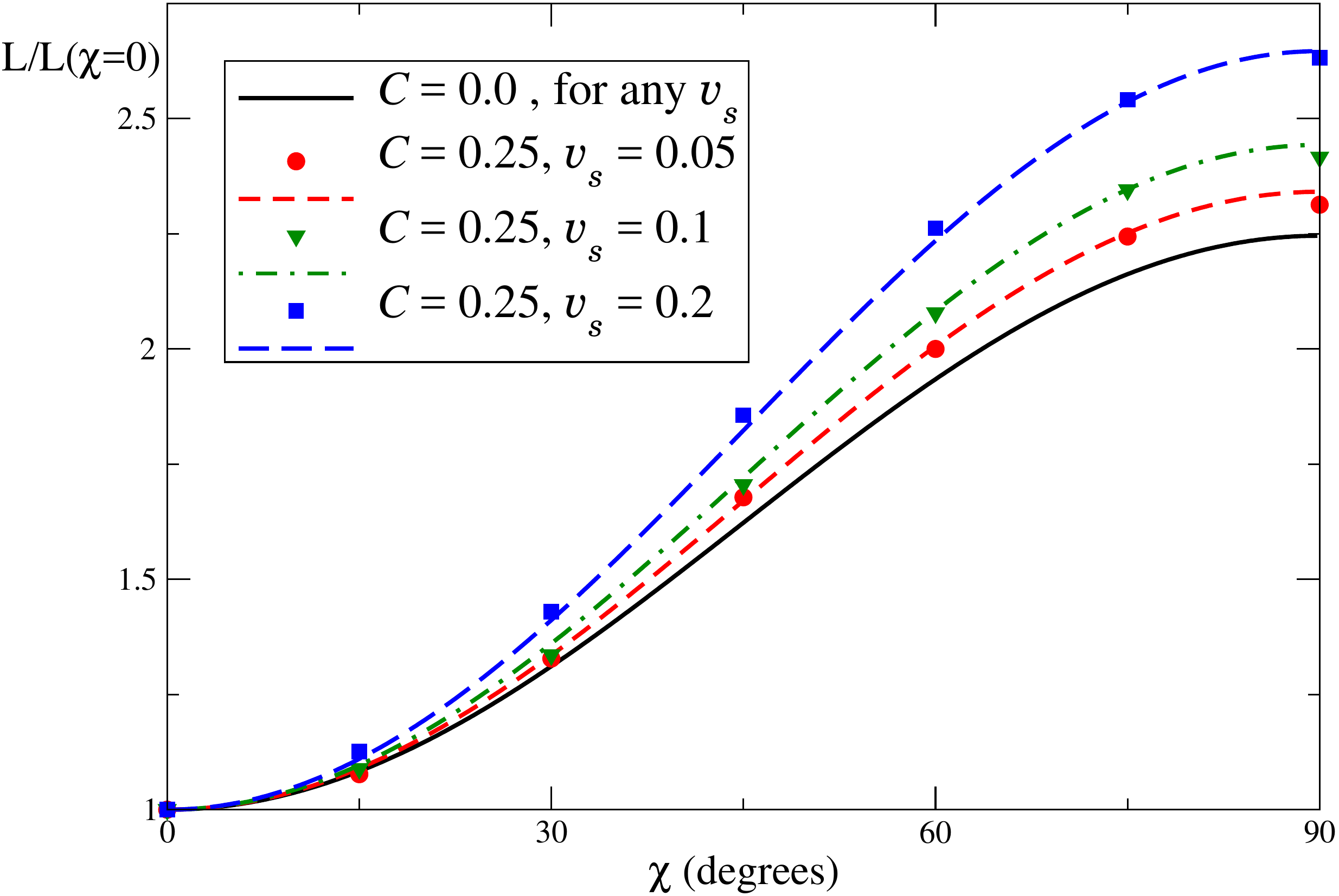}
  \caption{{\em Misaligned Rotator}. Dependence of the spin-down luminosity with inclination angle $\chi$. 
  Numerical data has been displayed along with relevant sections of expression \eqref{misaligned-fit}.
  The luminosity has been normalized to their aligned values, $L(\chi\text{=0})$; under this normalization, the results does not depend on the surface velocity $v_s$ for flat spacetime cases.
  }
 \label{fig:inclination} 
 \end{center}
\end{figure}

Figure \ref{fig:inclination} summarizes these results. Numerical data for $v_s = \left\lbrace 0.05, 0.1, 0.2 \right\rbrace $ (at fixed compactness $\mathcal{C}=0.25$) 
is displayed together with their corresponding curves taken from \eqref{misaligned-fit}. 
Notice for flat spacetime solutions, the normalized luminosity $L/L(\chi\text{=0})$ does not depend on the surface velocity; 
and thus, there is just a single curve describing the zero compactness cases.
Interestingly, the relation $ L \propto k_1 + k_2 \, \sin^2 \chi$ proposed by Spitkovsky in \cite{spitkovsky2006} fits very well the numerical data, although the coefficients $k_1$ and $k_2$ now depend on the compactness and the spin rate. 
A careful comparison shows that our results are in good agreement with those of Spitkovsky for the particular case $\mathcal{C}=0$ and $v_s=0.2$; 
which were later reproduced by many other authors within different frameworks (see fig.~6 of \cite{cerutti2017} and references therein).
Moreover, our results are also consistent with \cite{petri2015general}, which was the first to incorporate GR effects to the misaligned rotator.
Indeed, when fitting our solutions with $ L/L_o = a + b \, \sin^2 \chi$, we obtain the parameters listed on Table \ref{table2} for the cases $v_s = \left\lbrace 0.1, 0.2 \right\rbrace $, which are only slightly larger than to those found by Petri (Table 2 of \cite{petri2015general})\footnote{Notice that, apparently, there is a typo in one of the parameters listed on their Table 2 for the GR case with $R/R_{LC}=0.2$.}.

\begin{table}[!h]
\caption{ Best-fitting parameters $a \,/\, b$ for the inclination dependence of the luminosity, 
$L(\chi) / L_o = a + b \, \sin^2 (\chi) $.}
\vspace{0.2cm}
\centering {
\begin{tabular}{ c | c | c }         
$v_s$  & Flat ($\mathcal{C}=0.0$) & Kerr ($\mathcal{C}=0.25$) \\ \hline
 0.1   	&  $~ 0.984  \,/ \, 1.252 ~$  & $~ 1.130  \,/ \, 1.656 ~$   \\
 0.2    &  $~ 0.955  \,/ \, 1.170 ~$  & $~ 1.277  \,/ \, 2.056 ~$   \\
\end{tabular}
}
\label{table2}
\end{table}
%

\subsection{Braking Index}
\label{sec:break}

\begin{figure}
  \begin{center}
\includegraphics[scale=0.35]{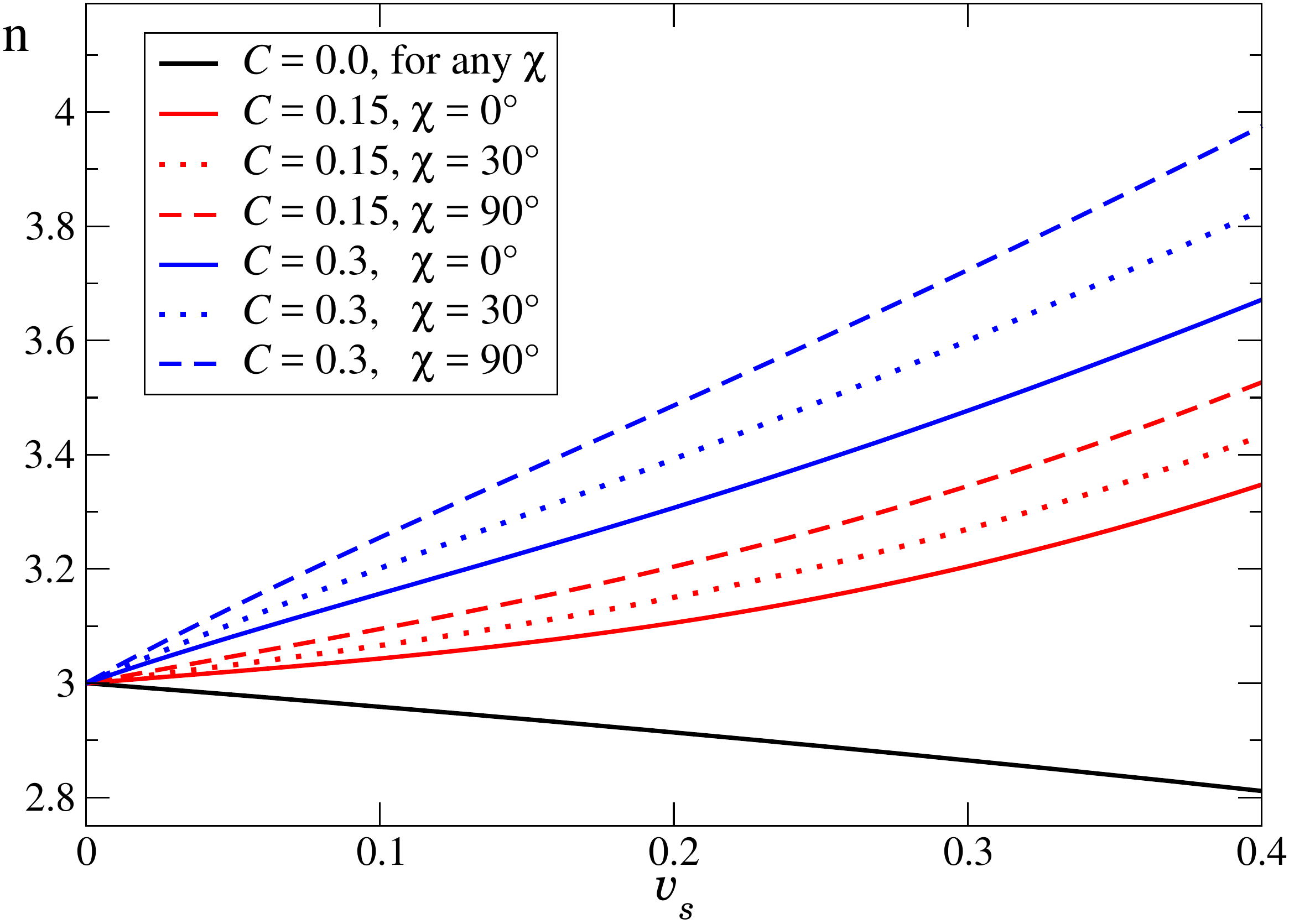}
\includegraphics[scale=0.35]{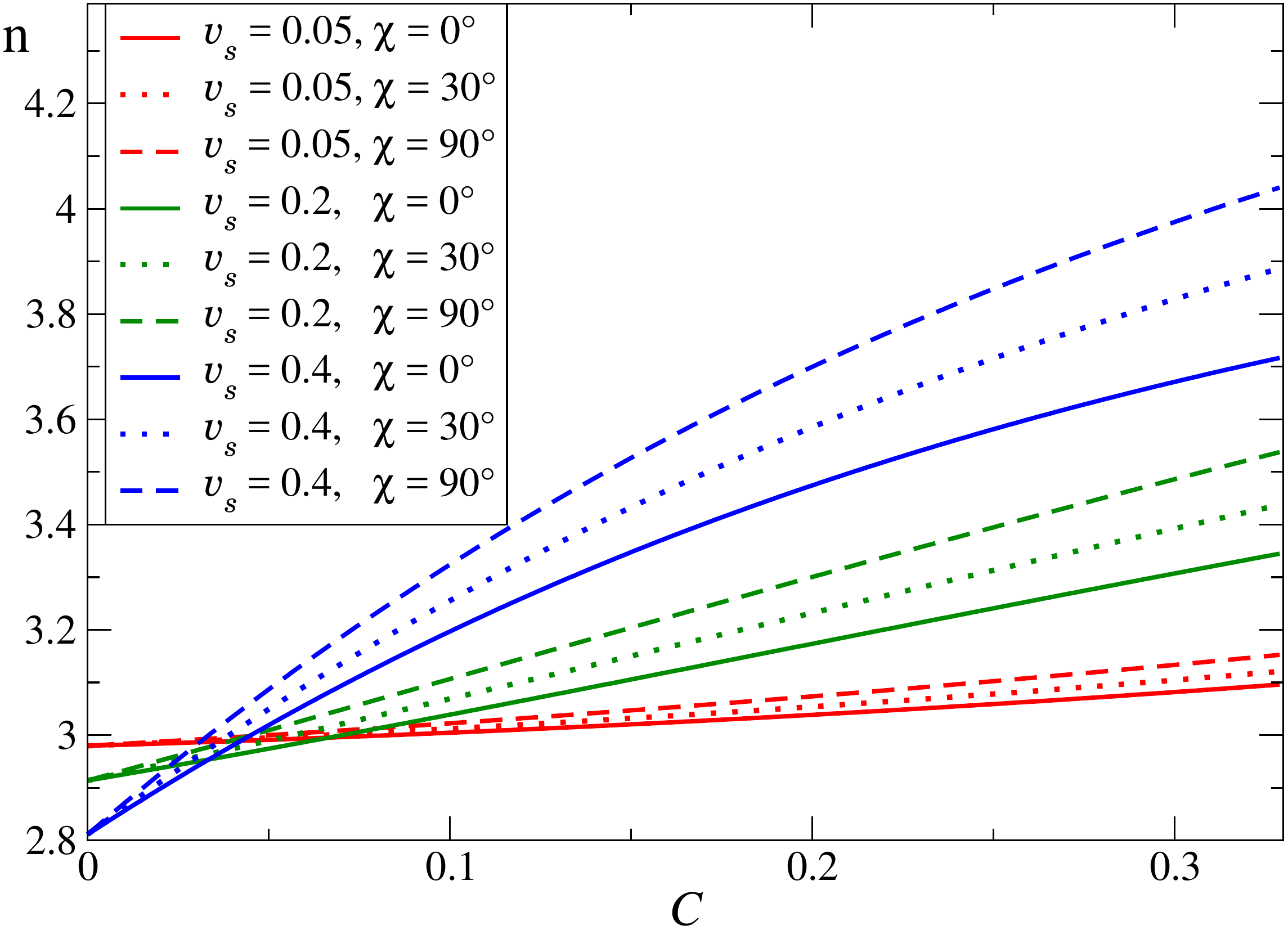}
  \caption{{\em Braking Index}. Estimated corrections to the usual point-dipole value, $n=3$. 
  \textbf{Top image:} dependence on surface velocity $v_s$, for some values of compactness and obliquity. 
  Notice that on the flat spacetime limit, the resulting braking index does not change with $\chi$.
  \textbf{Bottom image:} dependence on stellar compactness $\mathcal{C}$, for different spin rates and inclinations.
  }
 \label{fig:braking} 
 \end{center}
\end{figure}

One may rewrite the two fitting formulas \eqref{aligned-fit}-\eqref{misaligned-fit} from last sections into a single general expression,
\begin{equation}\label{gral-fit}
  L (\mathcal{C}, v_{s}, \chi)  =  L_o \, F(\mathcal{C}, v_{s}) \, \left[  1 +  k(\mathcal{C}, v_{s}) \, \sin^2 (\chi) \right]     
\end{equation}
where $L_o = \mu^2 \Omega^4 \equiv \left( \mu / R^2 \right) ^2 v_{s}^4$ and, 
\begin{eqnarray*}
k(\mathcal{C}, v_s ) &=& 1.24 + 8.14 \, v_s \, \mathcal{C}  \\
 F(\mathcal{C}, v_{s}) &=&  1.035 - 0.41  \, v_s + (0.16 + 3.9 \, v_s + 9.8 \, v_{s}^3) \, \mathcal{C} \\
 && + \, (11 \, v_s - 0.9) \, \mathcal{C}^2 \\
\end{eqnarray*}
From this approximate formula of the pulsar spin-down luminosity and equation \eqref{n}, one can readily obtain an estimated braking index $n$.
The calculation is rather straightforward and gives $n=n(\mathcal{C}, v_{s}, \chi)$ in a closed form.
However, the resulting expression is not very enlightening by itself and so, instead of writing it down, we illustrate the results by plotting 
some representative curves in Fig.~\ref{fig:braking} at  constant $\mathcal{C}$ and $v_s$ (top and bottom panels, respectively).  
These plots show a braking index ranging from $n\simeq2.8$ for a rapidly rotating star on flat spacetime, up to $n\simeq4$ for a rapidly rotating and highly compact NS. 
Actually, for realistic pulsars, $\mathcal{C} \gtrsim 0.1$ and $v_s \lesssim 0.2$, the braking index increases both with the compactness and the surface velocity and it would be in the range $n\simeq 3.0-3.5$.
Notice that, in the Newtonian regime, the braking index does not vary with the inclination angle $\chi$,
whereas in the relativistic case its value increases with misalignment.
We also note that, as expected, deviations from the standard value $n=3$ disappear for $v_s \ll 0.1$, which is the most common case for the observed pulsars. Unfortunately, the handful of pulsars with a reliable estimation of $n$ have periods of the order of $0.1-1$ s, for which the relativistic effects on the spin-down formula are negligible\footnote{There are many caveats regarding coherent timing analysis that complicate a comparison: 
possible occurrence of unseen glitches, shape of the residuals, choice of integration time, among others.}. 





\section{Conclusions}
 
We have presented a formalism and a numerical code to carefully analyze the force-free magnetospheres of neutron stars, by extending previous studies on black hole magnetospheres. We have performed several tests to show the correct implementation of the boundary conditions at the stellar surface, which is the main difference
with respect to the previous code. A careful and detailed decomposition of the eigenvectors of the evolution equations is required to apply suitable inner boundary conditions of our domain, which corresponds to a perfectly conducting neutron star.

We have studied, through three-dimensional time-dependent numerical simulations, general-relativistic neutron star magnetosphere within the force-free approximation. 
Our results confirm other recent numerical investigations, in particular regarding the total electromagnetic power radiated by the neutron star. 
By performing suitable fits of our results, we provide a quite generic formula for the luminosity of a rotating dipolar magnetic field as a function of the compactness of the neutron star, 
its angular velocity, and the misalignment angle between the spin and the magnetic dipolar moment.
From this formula, we have estimated deviations from the standard braking index value which will generally depend on these adimensional parameters.
These deviations are rather modest, leading to $n = 3.2 \pm 0.2$ for realistic millisecond pulsars.
Unfortunately, current astrophysical observations are not accurate enough to distinguish such small differences on the braking index. Even for the pulsars displaying an almost constant braking index which could be measured with high precision, the observed values are systematically below the standard $n=3$~\cite{pons2012pulsar, espinoza2016new}.
As it has been confirmed in this work, these deviations can not be associated to relativistic effects. Therefore, we can only conclude that, if the measurements are correct, there must be other physical mechanisms modifying the luminosity of a pulsar magnetosphere which are not captured by the simple force-free model.



\section{Acknowledgments}

We would like to thank Daniele Vigan\`{o} for several helpful discussions throughout the realization of this work.
F.C. and O.R. acknowledge financial support from Conicet, SeCyT-UNC and MinCyT-Argentina.
F.C. and C.P. acknowledge support from the Spanish Ministry of Economy and Competitiveness grants AYA2016-80289-P and AYA2017-82089-ERC (AEI/FEDER, UE). C.P. also acknowledges support from the Spanish Ministry of Education and Science through a Ramon y Cajal grant.
F.C. thanks Conicet \textit{financiamiento parcial de estadias breves en el exterior para becarios posdoctorales} program and the Departament de F\'i{}sica \& IAC3 for hosting his stay at the Universitat de les Illes Balears, where parts of this work were completed.
This work used computational resources from Mendieta Cluster (CCAD, Universidad Nacional de C´ordoba) and
Pirayu Cluster (supported by the ASACTEI, Gobierno de la Provincia de Santa Fe, Proyecto AC-00010-18), which are both part of the SNCAD -- MinCyT-Argentina.


\appendix

\section{Characteristic Decomposition and Boundary Operators}\label{sec:apxA}

In this appendix, we build appropriate boundary operators $ L^{(a)}(\cdot)$ (see eqn.~\eqref{boundary_op}) to deal with a perfectly conducting surface via penalty terms.  
The construction strongly relies on the characteristic structure of the particular evolution system under consideration.
Thus, before going to the main result concerning the full force-free system used in this paper, we shall first illustrate the method by applying it to vacuum electrodynamics
and to a simpler version of the force-free system.
The first one is the usual linear theory with transversal propagation modes, and serves as example for the whole construction.
While the second system, already nonlinear, posses a much simpler characteristic structure that helps to better understand the transition to the full theory. 
This simpler approach to force-free electrodynamics has been widely used in the literature, as in e.g. \cite{Palenzuela2010Mag}, and so it might be also interesting in its own right. 

These systems can be described by the same covariant equations,
\begin{eqnarray}
  \nabla_b F^{ab} &=& I^a  \label{Maxwell}\\
 \nabla_b F^{*ab} + \nabla^a \phi &=& \kappa n^a \phi \label{Faraday} 
\end{eqnarray}
where an extra dynamical field, $\phi$, has been added to handle the magnetic divergence-free constraint (e.g.~\cite{Dedner, Komissarov2004b, Mari}).
If $I^a = 0$ one recovers usual electrodynamics in vacuum. 
On the other hand, the presence of the plasma can be captured by an effective force-free current density, 
\begin{equation}\label{ohm-current}
 I^a = \rho \left( n^a + \frac{S^a}{B^2} \right) 
\end{equation}
which includes the drift current only. In this approach the force-free condition, $E \cdot B = 0$, needs to be enforced separately:
either by damping strategy like the one used on \cite{alic2012}, or by cutting electric field at each Runge-Kutta substep as in e.g.~\cite{Komissarov2004b, Palenzuela2010Mag}. 
Notice, $\rho:= \frac{1}{\sqrt{h}} \partial_k (\sqrt{h} E^k ) $, represents here the electric charge density. 
The resulting evolution equations are nonlinear and, as it can be shown, they constitute a strongly hyperbolic system. \\ 

Let us introduce some important notational conventions for the remaining of this section.
We shall denote the set of dynamical fields $U^{\mu}$ as,
\begin{equation}
 U^{\mu} = \left( \begin{array}{c} \phi \\ E^i \\ B^i \end{array}  \right) 
\end{equation}
(where the index $i$ is used to denote spatial vectors) and its ``dual'' element $\Theta_{\mu}$,
\begin{equation}
 \Theta_{\mu} = \left\lbrace  \phi,  E_i,  B_i \right\rbrace  
\end{equation}
Characteristic decompositions will be taken respect to the wave front propagation direction, given by a unit vector $m^i$
that will be identified in this context with the boundary surface normal. Some useful abbreviations are defined: 
\begin{equation}
 A_m \equiv m_i A^i \text{;} \quad A_{p}^i \equiv A^i - A_m \, m^i \text{;} \quad A_{q}^i \equiv m_k \, \epsilon^{kij} A_j  \nonumber
\end{equation}
for any vector $A^i$. Notice $A_{p}^i$ and $ A_{q}^i $ are orthogonal to each other, and both tangent to the boundary surface.

\subsection{Vacuum electrodynamics}\label{sec:Max_example}

The physical modes in Maxwell electrodynamics are transversal to the propagation direction $m^i$, and propagate at the speed of light with eigenvalues $\lambda^{\pm} = \beta_m \pm \alpha$. 
The associated characteristic basis (and co-basis) elements can be written as,
\begin{equation}
 U_{1}^{\pm} = \frac{1}{\sqrt{2}}\left( \begin{array}{c} 0 \\ \pm e_{p}^i \\ -e_{q}^i \end{array}  \right) \quad \text{;} \quad   
 U_{2}^{\pm} =  \frac{1}{\sqrt{2}}\left( \begin{array}{c} 0 \\ e_{q}^i \\ \pm e_{p}^i \end{array}  \right)\nonumber
\end{equation}
\begin{equation}
 \Theta_{1}^{\pm} = \frac{1}{\sqrt{2}}\left\lbrace  0, \pm e_{pi} , -e_{qi} \right\rbrace  \quad \text{;} \quad   
 \Theta_{2}^{\pm} =  \frac{1}{\sqrt{2}} \left\lbrace  0,  e_{qi} , \pm e_{pi}   \right\rbrace \nonumber
\end{equation}
being ($e_{p}^i$, $e_{q}^i$) any two orthogonal transversal directions. 
There are in this case three extra modes associated with constraints, playing no role in the construction of the boundary operators.
The operators $L^{(a)}(\cdot)$ will only involve here the physical modes, and shall be written as:
\begin{eqnarray*}
 && L_1 = \Theta^{+}_1 - R_{11} \Theta^{-}_1 - R_{12} \Theta^{-}_2 \\
 && L_2 = \Theta^{+}_2 - R_{21} \Theta^{-}_1 - R_{22} \Theta^{-}_2 
\end{eqnarray*}
The idea is now to find the coefficients, $R_{ij}$, as to remove from both $L_i$ the action on any free (i.e.~non-restricted) component of the boundary fields.
That means, in our case, to remove the components of $\Theta^{+}_1$ and $\Theta^{+}_2$ acting on the tangential magnetic field.
An immediate consequence of such requirement is that,
\begin{equation*}
 R_{11} = 1 \quad \text{;} \quad
 R_{12} = R_{21} = 0 \quad \text{;} \quad
 R_{22} = -1 
\end{equation*}
The action on the difference $dU^{\mu} = U^{\mu}_o - U^{\mu}$ then reads,
\begin{eqnarray}
 && L_1 (dU) = \sqrt{2} \, (dE \cdot e_p ) \\
 && L_2 (dU) = \sqrt{2} \, (dE \cdot e_q )
\end{eqnarray}
This way, we are able to control both tangential components of the electric field at the boundary, i.e.:
\begin{equation}
 E_{p}^{i} |_{\text{boundary}} = E^{i}_{op} 
\end{equation}

Notice the boundary is acting as a mirror, reflecting back part of the outgoing waves while keeping fixed the tangential components of the electric field.

\subsection{Alternative force-free system}

We look now at the evolution system which arises from a standard ($3+1$)-decomposition of equations \eqref{Maxwell}-\eqref{ohm-current}. 
In this system, the transversal subspace (with eigenvalues $\lambda^{\pm} = \beta_m \pm \alpha$) appear slightly modified and there is an extra physical mode\footnote{The 
presence of the plasma turns the characteristic mode associated to the constraint $\nabla \cdot E = 0$ in vacuum electrodynamics, into a new physical propagation linked to electric charge density.}, 
which might be incoming or outgoing according to the sign of $\lambda_E := \beta_m + \alpha \sigma_{E}$ (where $\sigma_{E}=\frac{S_m}{B^2}$). 
The complete eigen-system is here,
\begin{eqnarray}
&& U_{0}^{\pm} = \frac{1}{\sqrt{2}}\left( \begin{array}{c} 1 \\ 0 \\ \pm m^i \end{array}  \right)  \quad \text{;} \quad   
   U_{1}^{\pm} = \frac{1}{\sqrt{2}}\left( \begin{array}{c} 0 \\ \pm S_{p}^i \\ -S_{q}^i \end{array}  \right) \nonumber \\ 
&& U_{2}^{\pm} =  \frac{1}{\sqrt{2}}\left( \begin{array}{c} 0 \\ S_{q}^i \\ \pm S_{p}^i \end{array}  \right)   \quad \text{;} \quad
 U_{3} = \left( \begin{array}{c} 0 \\ b \, m^i - \sigma_{E} S_{p}^i \\ S_{q}^i \end{array}  \right)  \nonumber
\end{eqnarray}
\begin{eqnarray*}
 \Theta_{0}^{\pm} &=& \frac{1}{\sqrt{2}}\left\lbrace  1, 0 , \pm m_j \right\rbrace  \\
 \Theta_{1}^{\pm} &=& \frac{1}{\sqrt{2}} \left\lbrace 0, a_{\pm} S_{p}^2 m_j \pm S_{pj} , -S_{qj}  \right\rbrace  \\ 
 \Theta_{2}^{\pm} &=&  \frac{1}{\sqrt{2}}\left\lbrace 0, S_{qj} , \pm  S_{pj} \right\rbrace  \\ 
 \Theta_{3} &=& \frac{1}{b} \left\lbrace 0, m_j , 0  \right) 
\end{eqnarray*}
where we have defined, $ \text{  } a_{\pm} \equiv \frac{1}{b}(1\pm \sigma_{E}) \text{ ; }  b \equiv B^2 (1 - \sigma_{E}^2 )$.\\

\vspace{2mm}

\noindent \textbf{\underline{Outgoing case, $\lambda_E \leq 0$}.} Here there are two incoming physical modes, namely: $U_{1}^{+}$ and $U_{2}^{+}$. And thus, the general boundary operators
 \footnote{Constraint modes are related to the components $\phi$ and $B_m$ only, and decouple from the rest of the system as in the previous case.} 
 reads:
\begin{eqnarray*}
 && L_1 = \Theta^{+}_1 - R_{11} \Theta^{-}_1 - R_{12} \Theta^{-}_2 - R_{13} \Theta_3 \\
 && L_2 = \Theta^{+}_2 - R_{21} \Theta^{-}_1 - R_{22} \Theta^{-}_2 - R_{23} \Theta_3 
\end{eqnarray*}
The idea is, as before, to find the coefficients which removes from $L_i$ the action on any free component of the boundary fields.
That means, to remove the components of $\Theta^{+}_1$ and $\Theta^{+}_2$ acting on the tangential magnetic field.
As a consequence,
\begin{equation*}
 R_{11} = 1 \quad \text{;} \quad
 R_{12} = R_{21} = 0 \quad \text{;} \quad
 R_{22} = -1 
\end{equation*}
while it seems there is some freedom on the other coefficients $R_{13}$ and $R_{23}$. One possible choice being,
\begin{equation}
 R_{13} = \sqrt{2} \, S_{p}^2 \, \sigma_E  \quad \text{ ; } \quad R_{23} = 0 \nonumber
\end{equation} 
which results on no explicit enforcing of the normal electric field. 
The action would then read,
\begin{eqnarray}
&& L_1 (dU) = \sqrt{2} \, \frac{(dE \cdot S_{p})}{S_{p}^2} \label{ohm_p1_a}\\
&& L_2 (dU) = \sqrt{2} \, \frac{(dE \cdot S_{q})}{S_{p}^2}  \label{ohm_p1_b}
\end{eqnarray}

Another natural election would be to set $R_{13}=0$ and $R_{23}=0$. Obtaining, 
\begin{eqnarray}
&& L_1 (dU) = \sqrt{2} \,\left[  \frac{(dE \cdot S_{p})}{S_{p}^2} + \frac{\sigma_E}{b} dE_m \right]  \label{ohm_p2_a}\\  
&& L_2 (dU) = \sqrt{2} \, \frac{ (dE \cdot S_{q})}{S_{p}^2} \label{ohm_p2_b}
\end{eqnarray}

We shall use \eqref{ohm_p2_a}-\eqref{ohm_p2_b} for the general case in which the magnetic field is not tangential to the stellar surface,
so that the normal component of the electric field is enforced. While for the special case where $B_m = 0$,
we shall adopt \eqref{ohm_p1_a}-\eqref{ohm_p1_b} instead, fixing solely the tangential components.\\

\vspace{2mm}

\noindent \textbf{\underline{Incoming case, $\lambda_E > 0$}.} Now there are three incoming physical modes, namely: $U_{1}^{+}$, $U_{2}^{+}$ and $U_3$. 
 The general boundary operators are:
\begin{eqnarray*}
 && L_1 = \Theta^{+}_1 - R_{11} \Theta^{-}_1 - R_{12} \Theta^{-}_2  \\
 && L_2 = \Theta^{+}_2 - R_{21} \Theta^{-}_1 - R_{22} \Theta^{-}_2  \\
 && L_3 = \Theta_3 - R_{31} \Theta^{-}_1 - R_{32} \Theta^{-}_2 
\end{eqnarray*}

In the present situation, the restriction to the operators completely determine all the coefficients:
\begin{equation}
 R_{11} = 1 \text{;} \quad
 R_{12} = R_{21} = R_{31} = R_{32} = 0 \text{;} \quad
 R_{22} = -1 \nonumber
\end{equation}
from which one obtains,
\begin{eqnarray}
&& L_1 (dU) = \sqrt{2} \,\left[  \frac{(dE \cdot S_{p})}{S_{p}^2} + \frac{\sigma_E}{b} dE_m \right] \\ 
&& L_2 (dU) = \sqrt{2} \, \frac{ (dE \cdot S_{q})}{S_{p}^2} \\ 
&& L_3 (dU) = \frac{dE_m}{b} 
\end{eqnarray}
We observe here, as opposed to the case $\lambda_E \leq 0$, is not possible to enforce the tangential electric field alone. 
But recall this was required only for the case in which $B_m = 0$, and it can be shown that $\lambda_E = 0$ when the magnetic field is purely tangential. 
Hence, there are no inconsistencies in the scheme and everything fits together perfectly. 
It seems the extra condition for $E_m$, arising from the force-free constraint,  
is relevant to the characteristic structure of the theory and plays a crucial role in the construction of the boundary operators.

\subsection{Full force-free system}

\subsubsection{Characteristic Structure}

The fully nonlinear force-free system \eqref{evol:phi}-\eqref{evol:B} has two transversal modes ($\lambda^{\pm} = \beta_m \pm \alpha $) and 
two Alfv\'{e}n modes of eigenvalues $\lambda^{\pm}_{A} = \beta_m + \alpha \sigma^{\pm}_A $ ($\sigma^{\pm}_A = \frac{1}{B^2}(S_m \pm B_m \Delta)$). 
These are modes associated with physical propagation.
The remaining modes correspond to constraints: two of them are related with divergence cleaning propagations (of eigenvalues  $\lambda_{0}^{\pm} = \beta_m \pm \alpha $)
and the third one represents the algebraic force-free constraint with $\lambda_{\Psi} = \beta_m + \alpha \frac{S_m}{B^2} $.

We write here the full eigen-system:
\begin{eqnarray*}
&& U_{T}^{\pm} = \left( \begin{array}{c} 0 \\ A_{\pm}^i \\ C_{\pm}^i \end{array}  \right) \text{;} \quad   
 U_{A}^{\pm} =  \left( \begin{array}{c} 0 \\ \left[ 1-(\sigma^{\pm}_A )^2 \right] m^i - \sigma^{\pm}_A V_{p\pm}^i \\ V_{q\pm}^i \end{array}  \right) \\
&& U_{\Psi} = \left( \begin{array}{c} 0 \\ B^i \\ -E^i \end{array}  \right)   \text{;} \quad  
 U_{0}^{\pm} =  \left( \begin{array}{c} 1 \\ 0 \\ \pm m^i \end{array}  \right) + \frac{E_m}{A_{\pm}^2}\left( \begin{array}{c} 0 \\ - C_{\pm}^i \\ A_{\pm}^i \end{array}  \right) 
\end{eqnarray*}
with its associated co-basis,
\begin{eqnarray*}
&&  \Theta_{T}^{\pm} = \frac{1}{2A_{\pm}^2} \left\lbrace 0 \text{ , } A^{\pm}_j \mp E_m m_j  \text{ , } C^{\pm}_j \right\rbrace  \mp \frac{E_m B_m}{2A_{\pm}^2} \Theta_{\Psi} \\
&&  \Theta_{A}^{\pm} = \frac{1}{2 N_{A}^{\pm}} \left\lbrace \pm \frac{E_m}{\Delta} \text{ , } \frac{B_m}{\Delta^2} B_j -  m_j \pm  \frac{E_{qj}}{\Delta}  \text{ , } \frac{B_m}{\Delta^2} E_j \mp  \frac{B_{qj}}{\Delta} \right\rbrace \\
&&  \Theta_{\Psi} = \frac{1}{\Delta^2}  \left\lbrace 0 \text{ , } B_j \text{ , } E_j \right\rbrace  \\
&&  \Theta_{0}^{\pm}= \frac{1}{2} \left\lbrace 1 \text{ , } 0 \text{ , } \pm m_j \right\rbrace  \pm \frac{E_m}{2} \Theta_{\Psi}
\end{eqnarray*}
where we have defined:
\begin{eqnarray*}
 && A_{\pm}^i := B_{q}^i \mp E_{p}^i \text{,} \quad C_{\pm}^i := E_{q}^i \pm B_{p}^i \text{,} \\
 && V_{\pm}^i := \frac{1}{B^2} (S^i \pm \Delta B^i ) \text{,} \quad \Delta:=\sqrt{B^2 - E^2}
\end{eqnarray*}
and the normalization, $N_{A}^{\pm} := \left[(\sigma^{\pm}_A )^2 - 1\right] $.

\subsubsection{Eigenvalues Analysis}

To go further into the construction of the boundary operators it seems necessary to better understand the Alfv\'{e}n eigenvalues,
since they depend on the values that the background fields take at each point during the evolution. 
Hence, it is not obvious which of these modes will turn-out incoming or outgoing across the stellar surface. 

Assuming the perfect conducting condition \eqref{E-cond} at the boundary,
we were able to perform a very general (but rather straightforward) calculation which gives:
\begin{equation}
 \lambda^{\pm}_{A} = \frac{B_m}{B^2} \left\lbrace (v\cdot B) \pm \sqrt{ B ^2 (\alpha^2 - v^2 ) +  (v\cdot B)^2} \right\rbrace 
\end{equation}
where $v^i := \beta^i + \Omega \, \eta^i$. Under very mild assumptions like e.g. $\Omega R \lesssim 0.5$, $a \lesssim 0.5$ and $C \lesssim 0.8$, it is possible to infer:
\begin{equation}
 \alpha^2 - v^2 > 0 
\end{equation}
This means that there will be always an incoming and an outgoing Alfv\'{e}n mode at the stellar boundary, 
except in the special case when the magnetic field is tangential to the surface (i.e.~$B_m = 0$). 
Indeed, in such situations one has $\lambda_{\psi} = \lambda_{A}^{+} = \lambda_{A}^{-} = 0 $, 
which relates with the well known fact that Alfv\'{e}n waves cannot propagate orthogonal to the magnetic field.

\subsubsection{Boundary Operators}

According to the previous analysis, we would always have one incoming and one outgoing Alfv\'{e}n modes (i.e. $\lambda_{A}^{\pm} > 0$ and $\lambda_{A}^{\mp} < 0$); 
except in the special case where $B_m = 0$, where one has $\lambda_{A}^{+} = \lambda_{A}^{-} = 0 $. 
On the other hand, we also have incoming and outgoing transversal modes with eigenvalues $\lambda^{\pm} = \beta_m \pm \alpha$.
So there are always (provided $B_m \neq0$) two incoming physical modes that can be prescribed via penalties, and an extra unphysical $\Psi$-mode which may be incoming or outgoing
\footnote{The remaining modes, linked to dynamical enforcement of the constraint $\nabla \cdot B = 0$, will not be treated on the same footing. 
This differential constraint is handed with a different method to restrict possible incoming violations from the boundary \cite{FFE2}.}.


\vspace{2mm}

\noindent \textbf{\underline{Outgoing case, $\lambda_{\Psi} \leq 0$}.} There are two incoming modes: $U_{T}^{+}$ and $U_{A}^{\pm}$ (``$\pm$'' depending on the signs of  $\lambda_{A}^{\pm} $).
 The general boundary operators are: 
\begin{eqnarray*}
 && L_T = \Theta^{+}_T - R_{TT} \Theta^{-}_T - R^{\mp}_{TA} \Theta^{\mp}_A - R_{T\Psi} \Theta_{\Psi} - R_{T0} \Theta^{-}_0  \\
 && L^{\pm}_A = \Theta^{\pm}_A - R^{\mp}_{AA} \Theta^{\mp}_A - R_{AT} \Theta^{-}_T - R_{A\Psi} \Theta_{\Psi} - R_{A0} \Theta^{-}_0 
\end{eqnarray*}
The idea is to apply the operators to a generic field\footnote{It is very important to keep these fields separate from those background fields present in the co-basis elements $\Theta_i$. 
}, $\hat{U}$, and find the coefficients that cancel the terms containing $\hat{\phi}$ and $\hat{B}^i$.
For convenience, we shall consider $\hat{\Psi} \equiv \Theta_{\Psi}(\hat{U})$ instead of $\hat{B}_m$ on the calculations, 
by means of the relation: $E_m \hat{B}_m = \Delta^2 \hat{\Psi} - B_m \hat{E}_m - (E_p \cdot \hat{B}) - (B_p \cdot \hat{E}) $.\\

We shall solve first for $L_T$. After some redefinitions of the coefficients,
\begin{eqnarray*}
 && x_T := -\frac{A_{+}^2}{A_{-}^2} R_{TT} \quad \text{;} \quad 
 x^{\mp}_A := -\frac{A_{+}^2}{\Delta N_{A}^{\mp}  }  R^{\mp}_{TA} \\
 && x_{\Psi} := -2 A_{+}^2  R_{T\Psi}   \quad \text{;} \quad 
 x_0 := -\frac{A_{+}^2}{E_m} R_{T0}
\end{eqnarray*}
we obtain the following system:
\begin{eqnarray*}
 && \hat{\phi} ) \quad \, \, \, \, 0 = (x_0 \mp x^{\mp}_A ) E_m \\
 && \hat{B}^{i}_{\parallel}) \quad 0 = x_0 (E_p \cdot \hat{B}) \pm x^{\mp}_A (B_q \cdot \hat{B}) + (C_{+} + x_T \, C_{-})\cdot \hat{B} \\
 && \hat{\Psi} ) \quad \text{ } \, 0 = (x_T -1) E_m B_m  + x^{\mp}_A \Delta B_m  - x_0 (\Delta^2 + E_{m}^2 )  + x_{\Psi}
\end{eqnarray*}
which can be solved to get,
\begin{eqnarray}
&& x_T = (B_{p}^2 - E_{p}^2 + 2S_m )/A_{-}^2 \qquad\qquad \quad \text{ } x^{\mp}_A = \pm 2 E_m B_m /A_{-}^2  \nonumber \\
&& x_{\Psi} = 2 E_m B_m B^2 (1+\sigma^{\mp}_A - \sigma_{\Psi})/A_{-}^2 \quad \quad  x_0 = 2 E_m B_m /A_{-}^2 \nonumber 
\end{eqnarray}
and finally, 
\begin{equation}\label{L_T}
 L_T (dU) =  \frac{(A_{+}\cdot dE)}{A_{+}^2}  + \frac{E_m dE_m}{A_{+}^2 A_{-}^2} \left[ B_{m}^2 + \Delta |B_m | - E_{p}^2 \right]
\end{equation}\\

Analogously, we solve now for the operator $L^{\pm}_A$. 
\begin{eqnarray*}
 && x_T := -\frac{\Delta N_{A}^{\pm} }{A_{-}^2} R_{AT} \quad \text{;} \quad 
 x^{\mp}_A := -\frac{N_{A}^{\pm}}{N_{A}^{\mp}}  R^{\mp}_{AA} \\
 && x_{\Psi} := -2\Delta N_{A}^{\pm}  R_{A\Psi}   \quad \text{;} \quad 
 x_0 := -\frac{ \Delta N_{A}^{\pm}}{E_m} R_{A0} 
\end{eqnarray*}
and we obtain the following system:
\begin{eqnarray*}
 && \hat{\phi}) \quad  0 = \left[ x_0 \pm (1-x^{\mp}_A )\right] E_m \\
 && \hat{B}^{i}_{\parallel}) \, \, \, 0 = x_0 (E_p \cdot \hat{B}) \mp (1-x^{\mp}_A )(B_q \cdot \hat{B}) + x_T (C_{-}\cdot \hat{B}) \\
 && \hat{\Psi}) \quad  0 = (1 + x^{\mp}_A) \Delta B_m  + x_T E_m B_m -  x_0 (\Delta^2 + E_{m}^2 ) + x_{\Psi} 
\end{eqnarray*}
which can be solved to get,
\begin{equation}
 x_T = 0 \quad \text{;} \quad 
 x^{\mp}_A = 1 \quad  \text{;} \quad 
 x_{\Psi} = -2\Delta B_m \quad \text{;} \quad 
 x_0 = 0 \nonumber 
\end{equation}
Finally,
\begin{equation}\label{L_A}
 L^{\pm}_A (dU) = -\frac{dE_m}{N_{A}^{\pm}}
\end{equation}\\


\vspace{2mm}

\noindent \textbf{\underline{Special case, $B_m = 0$}.} In this spacial situation there is just one incoming mode, namely: $U_{T}^{+}$.
 The general boundary operator is:
\begin{eqnarray*}
 L_T &=& \Theta^{+}_T - R_{TT} \Theta^{-}_T - R^{+}_{TA} \Theta^{+}_A - R^{-}_{TA} \Theta^{-}_A \\ 
     & & -  R_{T\Psi} \Theta_{\Psi} - R_{T0} \Theta^{-}_0  
\end{eqnarray*}
Recall we only can prescribe the tangential electric field here, so we need to further remove the $\hat{E}_m$ component in the construction.
Notice it will be now possible, due to the presence of an extra Alfv\'{e}n mode in the right hand side. 
We shall define again the coefficients,
\begin{eqnarray*}
 && x_T := -\frac{A_{+}^2}{A_{-}^2} R_{TT} \quad \text{;} \quad 
 x^{\mp}_A := -\frac{A_{+}^2}{\Delta N_{A}^{\mp}  }  R^{\mp}_{TA} \\
 && x_{\Psi} := -2 A_{+}^2  R_{T\Psi}   \quad \text{;} \quad 
 x_0 := -\frac{A_{+}^2}{E_m} R_{T0} 
\end{eqnarray*}
but now we obtain instead the system:
\begin{eqnarray*}
 && \hat{\phi}) \quad \, 0 = (x_0 + x^{+}_A - x^{-}_A ) E_m\\
 && \hat{B}^{i}_{\parallel}) \,\,\, (x^{+}_A - x^{-}_A)  (B_q \cdot \hat{B}) = x_0 (E_p \cdot \hat{B}) + (C_{+} + x_T \, C_{-})\cdot \hat{B} \\
 && \hat{\Psi}) \quad \, 0 =  x_{\Psi} - x_0 (\Delta^2 + E_{m}^2 )\\ 
 && \hat{E}_m ) \, \, \, 0 = (x_T -1) E_m  - (x^{-}_A + x^{+}_A)\Delta
\end{eqnarray*}
which leads to $x_{\Psi} = x_0 = 0 $ and,
\begin{equation*}
x_T = (B_{p}^2 - E_{p}^2 + 2S_m )/A_{-}^2 \quad \text{;} \quad x^{+}_A = x^{-}_A = -\frac{E_m E_{p}^2}{\Delta A_{-}^2} 
\end{equation*}
The operator then takes the form:
\begin{equation}
 L_T (dU) = \frac{1}{A_{+}^2} (A_{+}\cdot dE)
\end{equation}
and we see it only penalizes tangential components of the electric field at the boundary.\\

\vspace{2mm}

\noindent \textbf{\underline{Incoming case, $\lambda_{\Psi} > 0$}.} Now there are three incoming modes, namely: $U_{T}^{+}$, $U_{A}^{\pm}$ and $U_{\Psi}$.
 The general boundary operators are:
\begin{eqnarray*}
 && L_T = \Theta^{+}_T - R_{TT} \Theta^{-}_T - R^{\mp}_{TA} \Theta^{\mp}_A - R_{T0} \Theta^{-}_0  \label{LT}\\
 && L^{\pm}_A = \Theta^{\pm}_A - R^{\mp}_{AA} \Theta^{\mp}_A - R_{AT} \Theta^{-}_T - R_{A0} \Theta^{-}_0 \label{LA}\\
 && L_{\Psi} = \Theta_{\Psi} - R_{\Psi T} \Theta^{-}_T - R^{\mp}_{\Psi A} \Theta^{\mp}_A  - R_{\Psi 0} \Theta^{-}_0
\end{eqnarray*}

It is important to notice that there is not enough terms at the right hand sides of the first two expressions above to remove all the components we need;
in particular, $\hat{\Psi}$. However, we have an extra incoming (constraint) mode that will allow us to enforce $\hat{\Psi} = 0$ through the penalty method.
That allow us to solve the system, and we get:
\begin{eqnarray}
 && L_T (dU) = \frac{(A_{+}\cdot dE)}{A_{+}^2}  + \frac{E_m dE_m}{A_{+}^2 A_{-}^2} \left[ B_{m}^2 + \Delta |B_m | - E_{p}^2 \right] \nonumber\\ 
 && \qquad \qquad \quad  - \frac{E_m B_m}{A_{+}^2 A_{-}^2} (B^2 + \Delta |B_m |) \, \Theta_{\Psi}(dU) \\
 && L^{\pm}_A (dU) =  -\frac{dE_m}{N_{A}^{\pm}} + \frac{B_m}{N_{A}^{\pm}} \, \Theta_{\Psi}(dU)  \\
 && L_{\Psi} (dU) = \Theta_{\Psi}(dU) = \frac{1}{\Delta^2} \left[ (B\cdot dE) + (E\cdot dB) \right] 
\end{eqnarray}

\subsubsection{Degeneracies}

There are only three cases to consider separately from the general one we have described so far.
The first two happen when one of the Alfv\'{e}n subspace degenerates with one of the magnetosonic (i.e.~$\lambda_A = \lambda^{\pm}$), 
which is possible when: $E_m = 0 \text{, } \,  E_{p}^i \perp B_{p}^i \text{, } \, E_{p}^2 = B_{p}^2$ and $\Delta \equiv |B_m |$.
The third situation arises when this happens for both Alfv\'{e}n modes at the same time, only possible if $E^i = 0 \text{ and  } B_{p}^i = 0$.\\

\vspace{2mm}
\noindent
\underline{$\lambda_A = \lambda^{+}$:} $\quad$
 The relevant co-basis elements are, 
 \begin{eqnarray}
&& \Theta_{T}^{\pm} = \frac{1}{2B_{p}^2}  \left\lbrace 0 \text{ , } B_{qj} \text{ , } \pm B_{pj} \right\rbrace \nonumber\\
&& \Theta_{A}^{+} = \frac{1}{2B_{p}^2 B_{m}^2} \left\lbrace 0 \text{ , } (B_{p}^2 - B_{m}^2 ) B_{pj} + 2 B_{p}^2 B_{m} m_j   \text{ , } B^2 B_{qj} \right\rbrace \nonumber\\
&& \Theta_{A}^{-} = \frac{B^2}{2B_{p}^2 B_{m}^2} \left\lbrace 0 \text{ , } B_{pj}  \text{ , } B_{qj} \right\rbrace \nonumber
\end{eqnarray}
The boundary operators are fairly easy to solve; the only non-vanishing coefficients are $R_{TT} = -1$ and $R_{AA} = 1$. It leads to,
\begin{eqnarray}
 && L_T (dU) = \frac{(B_{q}\cdot dE)}{B_{p}^2} \\
 && L^{+}_A (dU) = \left[ \frac{dE_m}{B_m} -  \frac{(B_{p}\cdot dE)}{B_{p}^2}\right] 
\end{eqnarray}
where the corresponding incoming directions are,
\begin{equation}
  U_{T} =  \left( \begin{array}{c} 0 \\ B_{q}^i \\ B_{p}^i \end{array}  \right) \quad \text{;} \quad  U_{A}^{+} =  \left( \begin{array}{c} 0 \\ -B_{p}^i \\ B_{q}^i \end{array}  \right) 
\end{equation}\\
\vspace{2mm}
\noindent
\underline{$\lambda_A = \lambda^{-}$:} $\quad$
  The relevant co-basis elements are,
 \begin{eqnarray}
&& \Theta_{T}^{\pm} = \frac{1}{2B_{p}^2}  \left\lbrace 0 \text{ , } B_{qj} \text{ , } \pm B_{pj} \right\rbrace \nonumber\\
&& \Theta_{A}^{+} = \frac{B^2}{2B_{p}^2 B_{m}^2} \left\lbrace 0 \text{ , } -B_{pj}  \text{ , } B_{qj} \right\rbrace  \nonumber\\
&& \Theta_{A}^{-} = \frac{1}{2B_{p}^2 B_{m}^2} \left\lbrace 0 \text{ , } (B_{m}^2 -B_{p}^2 ) B_{pj} - 2 B_{p}^2 B_{m} m_j   \text{ , } B^2 B_{qj} \right\rbrace \nonumber
\end{eqnarray}
The boundary operators are fairly easy to solve; the only non-vanishing coefficients are $R_{TT} = -1$ and $R_{AA} = 1$. It leads again to,
\begin{eqnarray}
 && L_T (dU) = \frac{(B_{q}\cdot dE)}{B_{p}^2} \\
 && L^{+}_A (dU) = \left[ \frac{dE_m}{B_m} -  \frac{(B_{p}\cdot dE)}{B_{p}^2}\right] 
\end{eqnarray}
but now the Alfv\'{e}n direction in the penalty is given instead by,
 \begin{equation}
 U_{A}^{+} =  \frac{1}{B^2} \left( \begin{array}{c} 0 \\ (B_{p}^2 - B_{m}^2 ) B_{p}^i + 2B_{p}^2 B_m  m^i \\ B^2 B_{q}^i \end{array}  \right)  
\end{equation}
with $\lambda_{A}^{+} = \beta_m +  \alpha (B_{m}^2 - B_{p}^2 )/B^2 $, being the positive eigenvalue.
Some remarks follows:
\begin{itemize}
 \item The special situation in which $B_m = 0$ does not need any treatment, since is not allowed under these degenerate conditions.
 \item It may also be included here the penalization to the constraint mode, $L_{\Psi} (dU) = \Theta_{\Psi} (dU)$,
 whenever $\lambda_{\Psi} > 0$. 
\end{itemize}

\vspace{2mm}
\noindent
\underline{$E^i = 0 \text{ and  } B_{p}^i = 0$:} $\quad$ In such case, the characteristic system reduces to vacuum electrodynamics. 
And thus, we adopt the boundary operators discussed in Sec.~\ref{sec:Max_example}.

\bibliographystyle{unsrt} 
\bibliography{FFE}


\end{document}